# Advancing Seasonal Prediction of Tropical Cyclone Activity with a Hybrid AI-Physics Climate Model

Gan Zhang[1*], Megha Rao[1], Janni Yuval[2], Ming Zhao[3]

[1] Department of Climate, Meteorology, and Atmospheric Sciences, University of Illinois at Urbana-Champaign

[2] Google Research

[3] Geophysical Fluid Dynamics Laboratory, National Oceanic and Atmospheric Administration

*Corresponding Author: Gan Zhang (gzhang13@illinois.edu)

Abstract

Machine learning (ML) models are successful with weather forecasting and have shown progress in climate simulations, yet leveraging them for useful climate predictions needs exploration. Here we show this feasibility using Neural General Circulation Model (NeuralGCM), a hybrid ML-physics atmospheric model developed by Google, for seasonal predictions of large-scale atmospheric variability and Northern Hemisphere tropical cyclone (TC) activity. Inspired by physical model studies, we simplify boundary conditions, assuming sea surface temperature (SST) and sea ice follow their climatological cycle but persist anomalies present at the initialization time. With such forcings, NeuralGCM can generate 100 simulation days in ~8 minutes with a single Graphics Processing Unit (GPU), while simulating realistic atmospheric circulation and TC climatology patterns. This configuration yields useful seasonal predictions (July–November) for the tropical atmosphere and various TC activity metrics. Notably, the predicted and observed TC frequency in the North Atlantic and East Pacific basins are significantly correlated during 1990–2023 (r=~0.7), suggesting prediction skill comparable to existing physical GCMs. Despite challenges associated with model resolution and simplified boundary forcings, the model-predicted interannual variations demonstrate significant correlations with the observation, including the sub-basin TC tracks ($p<0.1$) and basin-wide accumulated cyclone energy ($p<0.01$) of the North Atlantic and North Pacific basins. These findings highlight the promise of leveraging ML models with physical insights to model TC risks and deliver seamless weather-climate predictions.

## 1. Introduction

Machine learning (ML) models recently made breakthroughs in weather forecasting (e.g., Bi *et al* 2023, Keisler 2022, Kochkov *et al* 2024, Lam *et al* 2023, Pathak *et al* 2022, Price *et al* 2023). Trained with atmospheric reanalysis or physical model data, these ML models delivered successful forecasts up to two weeks of lead time with skills comparable to or better than conventional numeric weather forecast (NWP) models. With computational costs at a fraction of NWP models ($10^{-3}$ to $10^{-5}$), the new ML models unlocked opportunities for improving operational weather service (e.g., Lang *et al* 2024) and advancing fundamental understanding of atmospheric predictability (e.g., Vonich and Hakim 2024). Similar to the early development of NWP and climate models (Phillips 1956), the success of ML models in weather forecasting also inspired researchers to explore their potential applications in climate modeling (Bretherton *et al* 2022, Eyring *et al* 2024). Nonetheless, the feasibility of conducting successful climate simulations and society-relevant climate predictions with the new ML models remains to be explored.

Recent efforts in leveraging the new ML models for climate simulations focused on attaining stable long-term simulations and emulating atmosphere-ocean interactions. For example, Cresswell-Clay *et al* (2024) trained ML emulators for the atmosphere and the surface ocean separately and showed that linking the two emulators can generate stable atmospheric simulations of the current climate for over 1000 years. Other endeavors emphasized achieving stable long-term simulations by incorporating various physical constraints. Bonev *et al* (2023) achieved one-year stable rollouts with the Fourier neural operator (FNO) by replacing an unrealistic flat geometry with spheric geometries. Based on the spheric FNO, Wang *et al* (2024) used gridded atmosphere and ocean data to train separate emulators and link them during roll-outs. With a configuration of lagged ensemble forecasting, their linked emulators achieved skillful seasonal predictions of the

El Niño–Southern Oscillation (ENSO). Watt-Meyer *et al* (2024) introduced mass and moisture constraints to the spheric FNO framework and completed an 80-year historical simulation with realistic atmospheric variability. This set of simulations performed well with in-sample climate forcings but showed unrealistic responses to out-of-sample climate forcings (i.e., zero-shot learning), such as high levels of sea surface temperature (SST) and carbon dioxide. To overcome such limitations, Beucler *et al* (2024) proposed to incorporate the physical knowledge of subgrid processes, which helped ML emulators trained with physical model outputs better generalize across climate regimes.

Distinct from those ML emulators, Kochkov *et al* (2024) developed an ML-physics hybrid model (NeuralGCM) and achieved multi-decade, stable atmospheric simulations. This model contains an atmospheric dynamical core like the conventional GCMs but replaces the parameterized subgrid physics with ML substitutes. Compared to the existing ML climate emulators, NeuralGCM stands out with its structural similarity to the conventional NWP models and atmospheric GCMs that are grounded on physical principles. NeuralGCM is highly skillful for weather forecasting and can incorporate observed SSTs to simulate climate anomalies, such as simulating realistic tracks and numbers of TCs in the active 2020 Atlantic hurricane season Kochkov *et al* (2024). These traits make NeuralGCM a promising candidate for modeling extreme risks and developing a seamless weather-climate prediction system (Brunet *et al* 2010, Hoskins 2013), provided that NeuralGCM can be configured to deliver skillful climate predictions.

While the current version of NeuralGCM lacks the means to simulate boundary conditions (e.g., ocean and sea ice) and the support for the atmosphere-ocean coupling, previous studies with physical models suggest that simple assumptions of boundary forcings can help establish a performance baseline for atmospheric GCMs in seasonal prediction tasks. Specifically, Zhao *et al*

(2010) showed that assuming persistent sea surface temperature (SST) anomalies with a climatological seasonal cycle can help an atmospheric GCM skillfully predict tropical cyclone (TC) activity in the North Atlantic and the Northeastern Pacific. Chen and Lin (2013) suggested that the prediction skill improves when the atmospheric GCM is initialized with observed conditions instead of random conditions from climate simulations. The success of these early studies with physical models builds on the thermal inertia of the tropical ocean and the strong influences of tropical SST on the global atmosphere (e.g., Shukla 1998) and TC activity (e.g., Gray 1984). The exploratory work with atmospheric GCMs served as a stepping stone for the ensuing development of more advanced prediction systems (e.g., Delworth *et al* 2020, Vecchi *et al* 2014).

Inspired by the recent NeuralGCM development and the previous physical studies, this study explores the feasibility of leveraging NeuralGCM to deliver skillful seasonal climate prediction. We emphasize TC activity since these storms are a leading contributor to life losses and economic damages (World Meteorological Organization 2021) and often remain challenging for physical GCMs to simulate (Roberts *et al* 2020). This TC focus also helps us leverage proven concepts and knowledge in physical model development (Chen and Lin 2013, Zhao *et al* 2010). Overall, this effort establishes a performance baseline for future model development that seeks to extend our climate modeling capability and deliver societally valuable predictions (Emanuel *et al* 2012, Lemoine and Kapnick 2024).

## 2. Data and Methods

### 2.1 Observational Data

The fifth-generation ECMWF atmospheric reanalysis (ERA5) (Hersbach *et al* 2020) serves as the primary data for the training, configuration, and validation of NeuralGCM simulations. The gridded ERA5 is generated by a numeric weather forecast model that follows physical laws and ingests multi-sourced observational data (e.g., weather station and satellite data). The original grid spacing of ERA5 is approximately 0.25-degree and contains variables at pressure levels and the surface (e.g., SST and sea ice coverage). The ERA5 data from 1979–2017 and 1979–2019 is used to train the deterministic and the stochastic NeuralGCM, respectively (Kochkov *et al* 2024). Since the training is based on narrow time windows (≤ 5 days), NeuralGCM does not directly learn the seasonal evolution trajectories. To facilitate the configuration and validation of retrospective prediction experiments, we regrid the ERA5 data to match the grid of NeuralGCM. While many recent ML studies use TC tracks extracted from the ERA5 for model evaluation, we evaluate TC predictions using the International Best Track Archive for Climate Stewardship (IBTrACS) (Knapp *et al* 2010). This dataset includes a collection of hurricane information based on multi-sourced observations and expert quality control. The best track dataset is widely used in TC research and real-world risk modeling and is generally considered more trustworthy than the reanalysis datasets (e.g., ERA5) that struggle with representing intense hurricanes (Dulac *et al* 2024).

## 2.2 NeuralGCM and Hindcast Experiments

NeuralGCM includes a differentiable dynamical core that solves the governing equations of atmospheric dynamics and a neural network that parameterizes unresolved processes of atmospheric columns (Kochkov et al., 2024). We use the pre-trained, 1.4-degree version of NeuralGCM to balance the need to conduct ensemble predictions and simulate realistic TC activity. At this resolution, NeuralGCM has two models: a deterministic model and a stochastic model (Kochkov et al., 2024). The deterministic model was extensively evaluated and showed promise in simulating realistic TC activity in the test of the year 2020 (Kochkov et al., 2024). The stochastic configuration uses random seeds to generate space-time correlated Gaussian random fields for perturbing initial conditions and insert stochasticity into the neural network parameterization. These random fields are independent of each other and conceptually resemble the NWP techniques of perturbing the initial fields and the parameterized model physics (Kochkov et al. 2024). Since we use the NeuralGCM versions trained by Kochkov et al. (2024) without modifying the model architecture or parameterized physics, we provide a high-level technical description in Supplementary Material and encourage interested readers to consult Kochkov et al. (2024) for more details.

We conduct hindcast experiments using both deterministic and stochastic configurations to assess the potential sensitivity of TC activity to the learned model physics parameters. To generate ensemble predictions with the deterministic model, we introduce perturbations to the initial conditions using a Gaussian random field. This field, initialized from a random seed, is applied to the learned correction within the NeuralGCM's encoder. Specifically, the encoder interpolates ERA5 initial conditions to sigma levels and subsequently learns a correction to this interpolation. We then perturb this correction by multiplying it by a factor of (1+random_field_value), where random_field_value represents the value from the generated Gaussian random field with correlation length of 1000 km for the deterministic model. We use this perturbation strategy to initialize twenty-member ensemble simulations at 0 UTC on July 1 for each year from 1990 to 2023. We also generated additional simulations (e.g., 1979–1989) to facilitate comparisons with previous TC studies that used physical models (Supplementary Materials).

Inspired by the seasonal prediction experiments by Zhao et al. (2010) and Chen & Lin (2013), we use the climatological seasonal cycle and persistent anomalies of SST and sea ice to drive the NeuralGCM. Based on the autocorrelation of SST and sea ice, this configuration can approximate the evolution of tropical SST (Chen & Lin, 2013) and sea ice (Bushuk *et al* 2022) during July–November. When calculating the anomalies of SST and sea ice at the initialized time, we use the daily climatology of 1991–2020 that is resampled using the monthly data. To ensure the consistency among variables and the configurations described by Kochkov et al. (2024), the initial states of the SST, sea ice, and atmosphere are acquired from the ERA5. At later steps of the prediction experiment, we force NeuralGCM with the pre-calculated SST and sea ice fields, namely the sum of their daily climate values and anomalies at the initialization time. Therefore, all the information needed for long-range predictions is available near the initialization time. We run the predictions for approximately five months to cover much of the TC season of the Northern Hemisphere. We acknowledge the assumption of persistent anomalies has limitations and consider the prediction skill of our experiments as a lower bound on the attainable skills. The 1.4-degree versions of NeuralGCM with the simplified boundary forcings can finish 100 simulation days in ~8 minutes with a single GPU (Supplementary Table 1).

## 2.3 Post-Processing and Evaluation

The combination of the stochastic NeuralGCM and modified boundary conditions yields stable multi-month predictions in most cases. While the hindcasts with the deterministic physics are generally stable (~98.5%) in the tropics, about 10% of the simulations with the stochastic physics configuration show spurious small-scale waves (Supplementary Figures 1 and 2) associated with unrealistic convection and stratosphere features. These waves mostly appear in the tropics and violate the weak gradient constraint of the real-world atmosphere (Charney 1963, Sobel and Bretherton 2000). While fixes are being explored, this study proceeds by labeling the simulations with spurious waves using a check of tropical variability. Specifically, we calculate the standard deviations of 500-hPa geopotential height in the zonal direction and compare the metric between the initial and later prediction steps. If spurious small-scale waves develop, they will substantially increase the zonal variability and thus the instability metric. If this metric exceeds two times the initial metric values at any latitudes during the roll-outs, we flag the corresponding simulation as unstable and assign all the fields to climate values. The flagging is robust to small changes in the threshold as the spurious waves usually amplify quickly once appearing in the rollouts. As the Supplementary Materials will show, the skills of the deterministic and the stochastic hindcasts in predicting TC activity are comparable. Unless otherwise specified, the analyses and discussion in the main text focus on the hindcasts with the more stable configuration with deterministic model physics.

The prediction evaluation includes selected environmental variables and metrics of TC activity. The dynamical variables include the 500-hPa geopotential height, which characterizes the steering flow that affects TC tracks, and the 200-850 hPa vertical wind shear, which affects TC genesis and development. While evaluating convection-related variables is important, the NeuralGCM version used here does not include precipitation variables. A new version that can simulate realistic precipitation is under development (Yuval *et al* 2024). We apply the TempestExtreme package (Ullrich *et al* 2021) to track TCs in our retrospective prediction experiments. The tracking uses the vorticity-based method and does not impose any wind speed thresholds. We follow most parameter choices of the TC tracker used by Kochkov et al. (2024) who tuned the parameters such that the TC counts of the ERA5 at 0.25-degree resolution match the values at 1.4 degrees. To better match the TC counts in the IBTrACS, we lower the vorticity threshold to $4 \times 10^{-5} s^{-1}$ and set the storm duration threshold to 54 hours.

Following previous studies of TC activity (e.g., Chen and Lin 2013, Zhang *et al* 2021, Zhao *et al* 2010), we mainly evaluate the ensemble mean and examine the metrics of anomaly correlation coefficient and root-mean-squared error. The evaluated variables include detrended environmental variables, regional TC counts, and the accumulated cyclone energy (ACE). The ACE is defined as the sum of the squares of the maximum wind speed (knots) of all the available track data with a scaling factor of $10^{-4}$. We also provide results from other models (e.g., Chen and Lin 2013, Johnson *et al* 2019, Zhang *et al* 2019) for reference. These models, such as the ECMWF seasonal forecasts (SEAS5) (Johnson *et al* 2019), are physical models with higher spatial resolutions of the atmosphere (e.g., 36-km grid spacing). We briefly discuss the performance of models in Section 3 and present additional analyses (e.g., SEAS5) and considerations for more comprehensive comparisons in Supplementary Materials.

## 3. Results

### 3. 1 Model Skill with Large-scale Atmospheric Environment

The NeuralGCM hindcasts with simplified boundary forcings simulate the atmospheric climate and seasonal cycle realistically (Figure 1). The July-November means of the 500-hPa geopotential height of the NeuralGCM and the ERA5 show consistent climate patterns. Their differences are the smallest in the tropics and the largest in the Arctic region. An inspection of the seasonal evolution of the zonal means of the 500-hPa geopotential height suggests the model biases grow over time. The initial biases emerge in the polar region and develop relatively rapidly during the transition season. The biases in the tropics are relatively small and comparable to those in a fully coupled physical prediction system (Supplementary Figure 3). Preliminary analyses (not shown) suggest that the high-latitude biases are related to the simplified boundary forcings, especially the ice representation in polar regions. The seasonally evolving geopotential biases affect the midlatitude jet streams and may ultimately distort some aspects of the tropical-extratropical teleconnections and TC activity (Wang *et al* 2020, Zhang *et al* 2016). Comparing other atmospheric variables suggests similar evolution between the climate states of the NeuralGCM and the ERA5 (not shown). Similar biases and consistency are present in the hindcasts with the stochastic version of NeuralGCM (not shown). Their overall consistency between the model climate and the observation is notable considering the simplified boundary forcings and the lack of complex atmosphere-land-ocean coupling in the NeuralGCM hindcasts.

The NeuralGCM hindcasts also show skills in predicting year-to-year variability of the monthly mean atmospheric environment. We examine variables including the 500-hPa geopotential, surface pressure, 1000-hPa temperature, and vertical shear of zonal wind (200-850 hPa) and find that the NeuralGCM hindcasts show various skill levels across the examined month leads (Supplementary Figures 4–7). Preliminary comparisons between the NeuralGCM hindcasts (Supplementary Figures 4–7) and operational seasonal predictions by a fully coupled physical model (Johnson et al 2019; Supplementary Figures 8–10) suggests the anomaly correlation coefficients with the observation are overall lower for the NeuralGCM hindcasts with simplified boundary forcings. Nonetheless, the anomaly correlation coefficients for NeuralGCM hindcasts show spatial-temporal patterns similar to those of the physical model. The anomaly correlation coefficients of the initial month are the highest and decrease with forecast lead time. While the decay quickly makes extratropical predictions unskillful, the prediction skill persists at much longer forecast lead in the tropics, as suggested by the relatively high correlation coefficients and low prediction errors (Supplementary Figures 4–7). The relatively high skill in the tropics corresponds to regions where the SST strongly regulates atmospheric variability (Shukla 1998), consistent with similar experiments with physical GCMs (Chen and Lin 2013).

We next focus on the prediction of the atmospheric environment in the Main Development Regions (MDRs). The MDRs, as outlined in Figure 1c, span over the tropical North Pacific and North Atlantic, which contribute a majority of TCs that form in the Northern Hemisphere (Goldenberg *et al* 2001, Doi *et al* 2013, Jien *et al* 2015, Zhang and Wang 2015, Feng *et al* 2021). Figure 2 shows the skill of the NeuralGCM with simplified boundary forcings in predicting the MDR atmospheric environment. The predictions of near-surface air temperature and the 500-hPa geopotential show the highest anomaly correlation coefficients with the observation across the three examined MDRs. For the prediction of these two variables, the anomaly correlation coefficients are statistically significant for each calendar month (Figure 2a–c). In comparison, the anomaly correlation coefficients for the surface pressure and the zonal wind shear are much lower but can remain statistically significant in August (Lead Days = 31-62). The anomaly correlation coefficients generally exceed those of persisting the monthly anomalies of June (Supplementary Figure 11). The findings thus suggest that the NeuralGCM with simplified boundary forcings can predict some aspects of the large-scale atmospheric variability in the MDRs. Such skills in predicting the large-scale environment are essential for the subseasonal-to-seasonal predictions of TC activity.

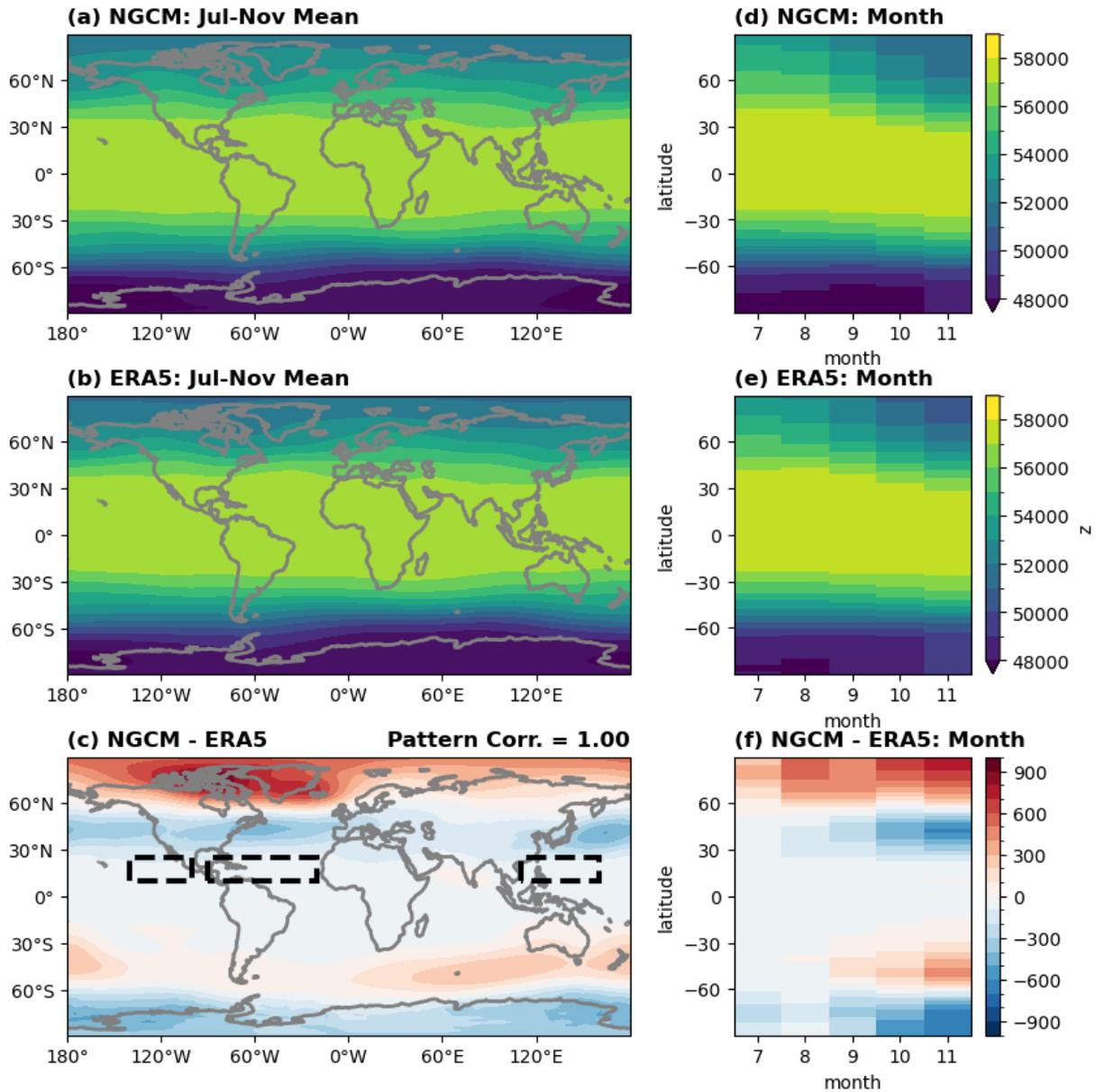

Figure 1 Comparison of the large-scale atmospheric environment in NeuralGCM hindcasts and the ERA5. (a) July-November ensemble mean (1990-2019) of the 500-hPa geopotential ($m^2\ s^{-2}$) in the NeuralGCM hindcasts. (b) Same as (a), but for the ERA5. (c) The difference between (a) the NeuralGCM hindcasts and (b) the ERA5. The pattern correlation (rounded to 2 decimal places) between (a) and (b) is denoted in the top right corner of (c). (d) The monthly evolution of the zonal mean of the 500-hPa geopotential ($m^2\ s^{-2}$) in the NeuralGCM hindcasts. (e) Same as (d), but for the ERA5. (f) The difference between (d) the NeuralGCM hindcasts and (e) the ERA5. The dashed black lines in (c) highlight the Main Development Regions (MDRs) where the environmental prediction skill is evaluated. The definitions of the MDRs roughly follow previous studies but use slightly larger latitudinal ranges (10°N–25°N). The longitudinal ranges for the MDRs of the North Atlantic, the Northeast Pacific, and the Northwest Pacific are 90°W–20°W, 140°W–100°W, and 110°E–160°E.

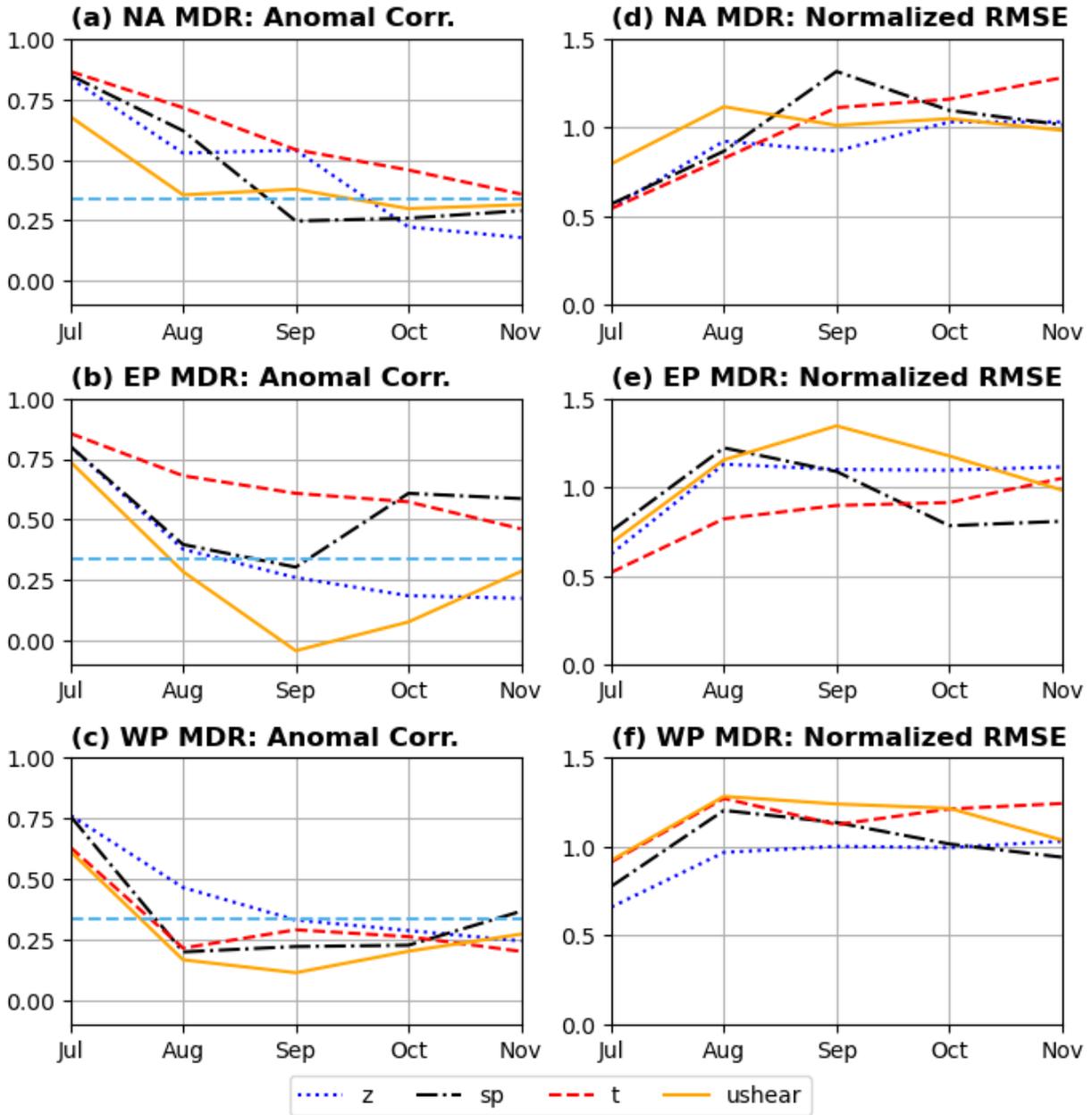

*Figure 2 Prediction skills of NeuralGCM for environmental fields in key TC regions. The plot shows selected atmospheric variables, including the 500-hPa geopotential (z, unit: $m^2\ s^{-2}$), surface pressure (sp, unit: hPa), 1000-hPa temperature (t, unit: K), and vertical shear of zonal wind (200-850 hPa) (ushear). (a) The anomaly correlation coefficient of the MDR in the North Atlantic (NA). (b) Same as (a), but for the Northeast Pacific (EP). (c) Same as (a), but for the Northwest Pacific*

*(WP). (d, e, f) Same as (a, b, c), but for the normalized root-mean-squared error. The scaling factor is the year-to-year standard deviations of the ERA5 data. The skill metrics are calculated using data from 1990-2023 for each grid point and then averaged over the MDR domains highlighted in Figure 1c. The input variables were detrended using a linear least-squared fitting. The 95% confidence threshold for the anomaly correlation coefficient is decided with t-statistics and denoted with horizontal cyan dashed lines in (a, b, c).*

## 3.2 Model Skill with TC Activity

The NeuralGCM hindcast can simulate a realistic spatial-temporal distribution of TC activity (Figure 3). The kernel density estimation of the simulated and observed TC activity shows consistent spatial patterns, including the high density of TC tracks in parts of the Northeast Pacific and the Northwest Pacific (Figures 3a-c). The relative track density among the Northern Hemisphere basins is also realistic, free of the common bias of many physical GCMs in severely underestimating TC activity in the Northern Hemisphere (e.g., Roberts et al. 2020). The seasonal cycles of TC activity in the Neural GCM and the observation are also similar (Figures 3d and 3e). The similarities include the latitudinal shift towards lower latitudes in the late season and the high concentration (~75%) of samples in July-September. The comparisons also show subtle biases of the NeuralGCM hindcasts with deterministic physics. For instance, the kernel density of TC tracks is too high near 10–20ºN in the Northeast Pacific; the decay of TC activity in the late season is also too fast, with October-November accounting for 17% instead of 25% of tracks. Similar seasonality biases are present in the NeuralGCM hindcasts with stochastic physics (not shown). This suggests these biases might arise from the simplified boundary forcings and biases in simulating the large-scale environment (Figure 1f), though the parameter choice of tracking algorithms may also be a contributing factor.

The NeuralGCM hindcast also simulates interannual variations of TC activity that are significantly correlated with the observation. The correlations between the seasonal prediction and the observation of the basin-wide TC frequency are statistically significant in the North Atlantic and the Northeast Pacific (Figure 4). Despite the much lower model resolution and computational costs, the NeuralGCM hindcasts demonstrate skill comparable to previous physical model simulations with similar simplified boundary forcings (Chen and Lin 2013; Zhao et al 2010) or a more realistic representation of boundary forcings (e.g., Zhang et al 2019) in those two basins . Table 1 shows a direct, like-for-like comparison of prediction skill for TC frequency. When model performance is ranked for the North Atlantic and the Northeast Pacific, the NeuralGCM hindcasts with deterministic physics are comparable to or better than at least one of the examined physical models. The prediction skill is associated with the relationship between TC activity and the large-scale atmospheric environment; moreover, the NeruralGCM hindcast can also predict at least some aspects of TC activity on the subseasonal scale, in hyperactive seasons, and beyond the model training period (Supplementary Materials).

*Table 1 Correlation of basin-wide TC frequency between seasonal predictions and the best track observations. The NeuralGCM results contain two rows that represent the hindcasts with the deterministic version (top) and the hindcasts with the stochastic version (bottom), respectively. The value ranges of NeuralGCM indicate 95%-confidence level intervals estimated using resampling with replacement. Except for the smaller number of resampling runs (N=1000), the*

*other settings of the skill estimation are similar to those in Zhang et al. (2019). The ensemble size and evaluation period are consistent between studies so the comparisons are relatively fair.*

| Reference and Evaluation Configuration | Model | N Atlantic | NE Pacific | NW Pacific |
|---|---|---|---|---|
| Chen and Lin (2013) 5-member ensemble 1990–2010 | HiRAM | 0.88 | 0.61 | 0.34 |
| | NeuralGCM | 0.74–0.87 0.69–0.89 | 0.54–0.73 0.37–0.71 | 0.19–0.47 0.05–0.37 |
| Zhang et al. (2019) 12-member ensemble 1981–2014 | FLOR | 0.60–0.75 | 0.47–0.60 | 0.27–0.44 |
| | NeuralGCM | 0.67–0.76 0.63–0.78 | 0.53–0.65 0.41–0.60 | 0.09–0.23 0.01–0.17 |

The prediction of TC frequency in the Northwest Pacific and the North Indian Ocean is relatively poor (Figures 4c and 4d) and appears related to the prediction skill of large-scale environment. Consistent with the results of the basin-wide TC frequency, the predicted track density in open-ocean areas is significantly correlated with the observation in parts of the North Atlantic and North Pacific but not the North Indian Ocean (Figure 5a). The regions with high prediction skills are consistent with physical model simulations and predictability analysis (Zhang et al 2019). Comparable basin-wide and regional correlations for the NeuralGCM hindcasts with the stochastic configuration (Figure 5b and Supplementary Figure 12). Since the environmental constraint of convective activity is crucial for long-range predictions (e.g., Shukla 1998), low skill in predicting TC activity is likely associated with large-scale environmental variables. For instance, the prediction skill of the atmospheric environment of the MDR of the Northwest Pacific is lower than that of the North Atlantic and the Northeast Pacific (Figure 2). In the North Indian Ocean, the prediction skill of local environmental variables is low (e.g., Supplementary Figure 7), and the TC-environment relationship is weak, making skillful seasonal predictions challenging (Supplementary Materials).

Interestingly, the NeuralGCM hindcast and the observation show significant correlations in the accumulated cyclone energy (ACE; Section 2.3). For the experiments with deterministic physics, we identified statistically significant correlations for the North Atlantic ($r = 0.68$), the Northeast Pacific ($r = 0.59$), and the Northwest Pacific ($r = 0.43$) (Figure 4). The hindcast with the stochastic physics shows comparable or better skill in predicting the ACE of the North Atlantic ($r = 0.69$), the Northeast Pacific ($r = 0.52$), and the Northwest Pacific ($r = 0.57$) (Supplementary Figure 12). We also compared the skill of these hindcast experiments and a physical model with a higher spatial resolution (Vecchi et al. 2014; Zhang et al 2019). During the 1981-2014 period, the NeuralGCM hindcasts and the physical model have comparable correlation coefficients in predicting the ACE in the North Atlantic, but the skill of the NeuralGCM hindcasts is notably lower with the Northeast and Northwest Pacific (not shown). The skill difference in predicting the ACE is possibly attributable to regional model biases and difficulties of the NeuralGCM hindcasts in representing intense TCs (>60 m s$^{-1}$) (Supplementary Figure 13). Nonetheless, this intensity-related issue is expected considering the lower resolution of the NeuralGCM (1.4-degree vs ~0.5-degree) used to generate our hindcasts.

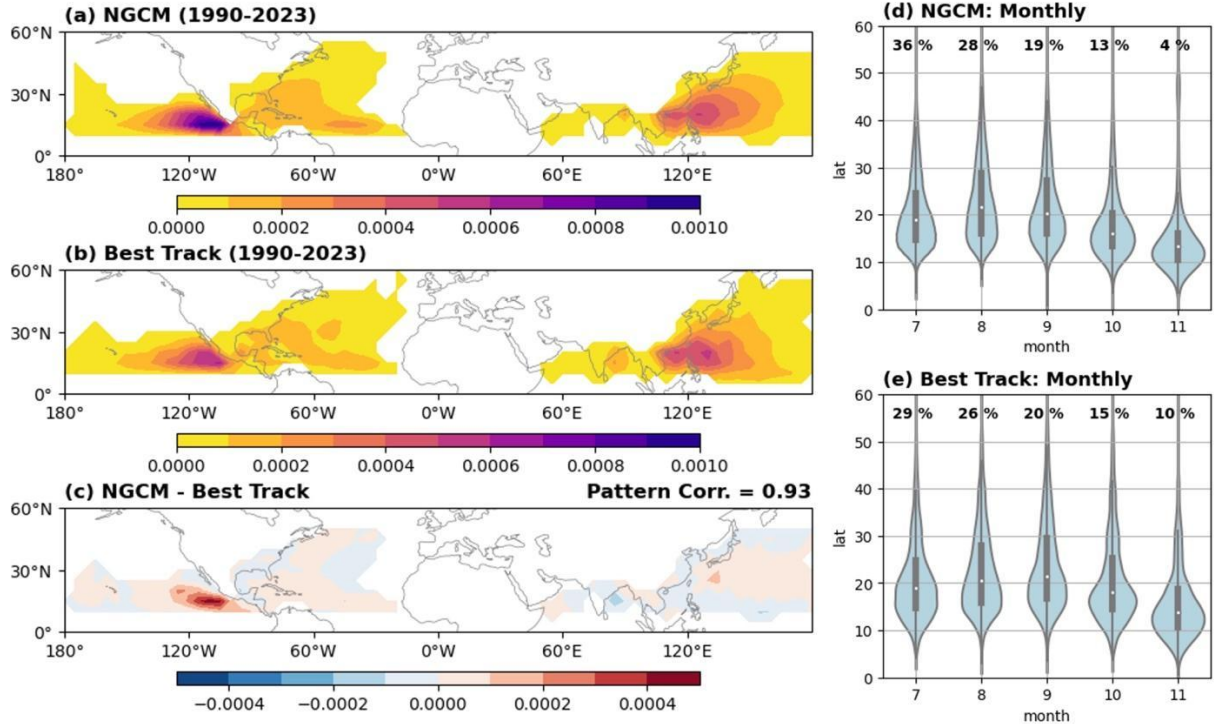

Figure 3 Comparison of July-November TC activity in NeuralGCM hindcasts and the IBTrACS (1990-2023). (a) Kernel density estimates (KDEs) of the TCs identified in the NeuralGCM hindcasts. (b) Same as (a), but for the best track observations. (c) The difference between (a) the NeuralGCM hindcasts and (b) the IBTrACS. The KDEs are calculated on a latitude-longitude grid with 5-degree spacing, and the unit of KDEs is $1 / (25 \text{ degrees}^2)$. The pattern correlation between (a) and (b) is denoted in the top right corner of (c). (d) The monthly evolution of the latitudinal distributions of the Northern Hemisphere TCs in the NeuralGCM hindcasts. The miniature box-and-whisker plots show the median, the interquartile range (IQR), and the 1.5×IQR range. The monthly percentage of the track points in the July-November total is denoted in the upper part. (e) Same as (d), but for the best track.

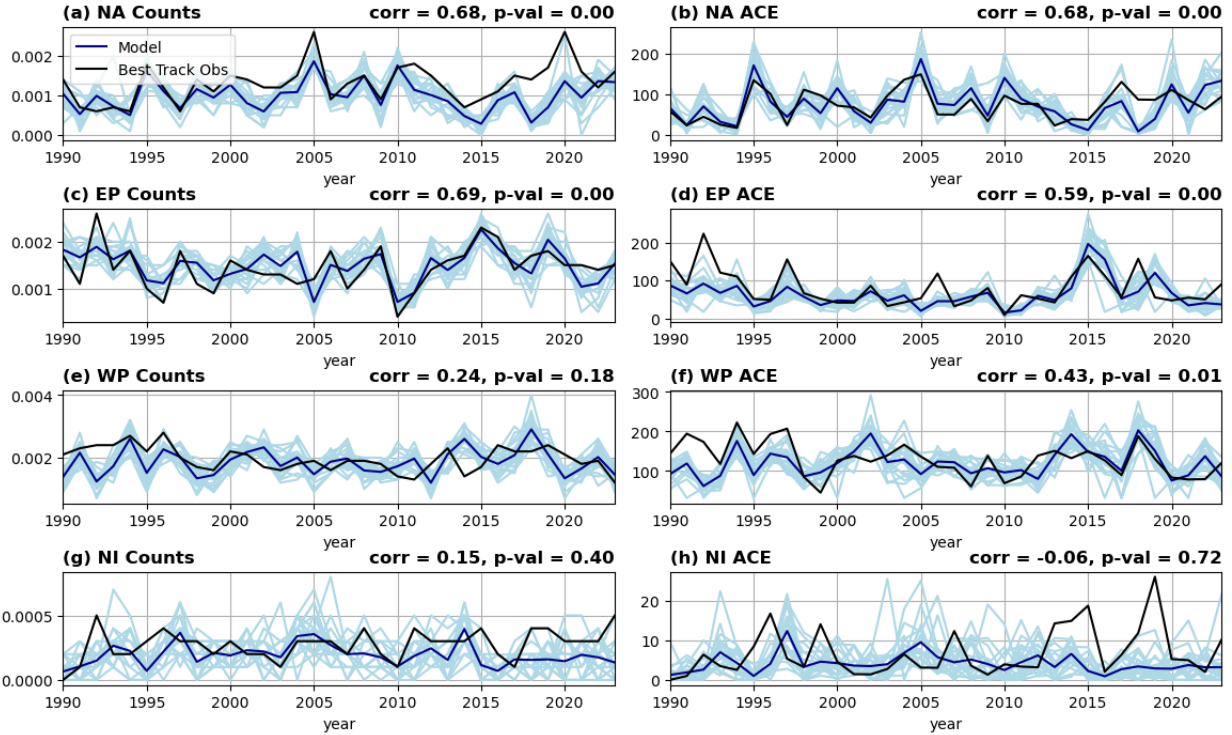

*Figure 4 Predictions and observations of the basin-wide TC counts and the accumulated cyclone energy (ACE; unit: $10^{-4}$ knot$^2$) (1990-2023). (a) North Atlantic TC counts, (b) North Atlantic ACE. (c–d), (e–f), and (g–h) Same as (a–b), but for Northwest Pacific, Northeast Pacific, and North Indian Ocean, respectively. The black line shows the annual hurricane counts in the observation (July–November). The blue lines show the counts in the NeuralGCM hindcasts. The dark blue line shows the ensemble mean, and the light blue line shows the individual ensemble members. The correlations between the ensemble means and the observation, as well as the associated statistical significance, are denoted in the top right corner of each subplot.*

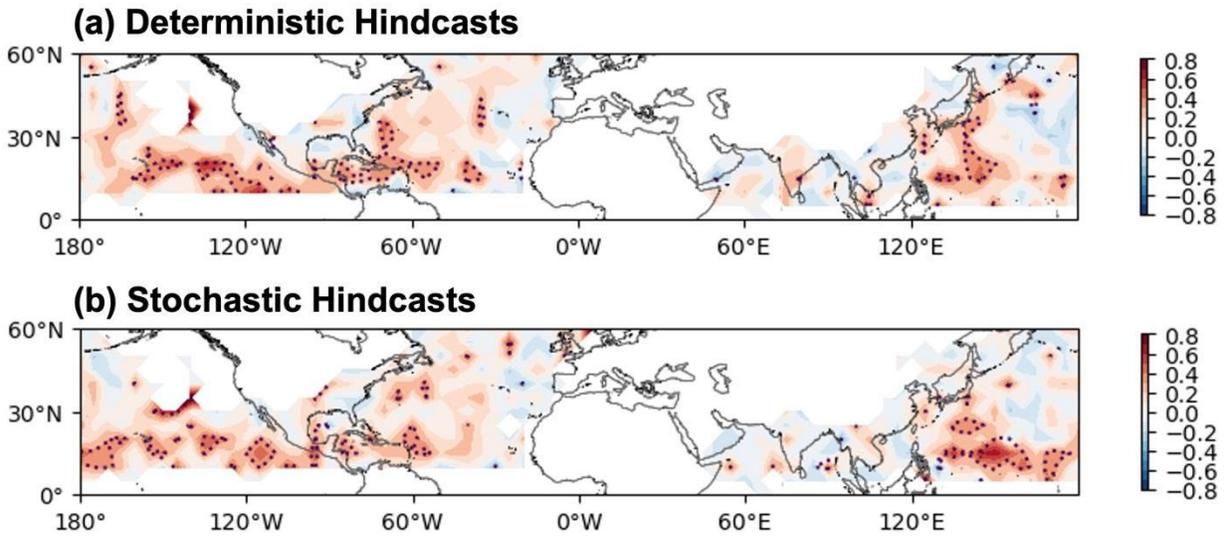

*Figure 5 The correlations between the cell counts of TC tracks in the NeuralGCM hindcast and the best track observation. (a) Hindcasts with deterministic physics. (b) Hindcasts with stochastic physics. The annual counts (July-November) are calculated based on a 5-degree grid. The dashed contours highlight the correlation coefficients at the 90% confidence level.*

## 4. Summary and Discussion

This study conducts experimental seasonal predictions with the newly available NeuralGCM and simplified boundary forcings. Inspired by earlier studies with physical GCMs, the hindcast experiments focus on July to November which account for most TC activity in the Northern Hemisphere. The NeuralGCM hindcasts of 1990–2023 can simulate realistic climate states of the atmosphere environment and TC activity. When predicting atmospheric variability, the NeuralGCM hindcast shows statistically significant anomaly correlation coefficients with the ERA5 reference at various forecast leads. Consistent with experiments conducted with physical GCMs, the skills are the highest for the environmental variables in the tropics. The hindcast also achieves relatively high skills in predicting seasonal metrics of TC activity, notably in the North Atlantic and the Northeast Pacific (r=~0.7). For instance, the prediction skill of TC activity metrics such as basin-wide TC frequency is comparable to physical models with much higher spatial resolution (e.g., Chen and Lin 2013) or more complex coupled processes (e.g., Zhang et al 2019) (Table 1). Despite challenges associated with intense TCs and some aspects of regional activity (e.g., the North Indian Ocean), the model-predicted interannual variations show significant correlations with the observation, including the sub-basin TC tracks ($p<0.1$) (Figure 5) and basin-wide accumulated cyclone energy ($p<0.01$) of the North Atlantic and North Pacific basins (Figure 4). The skill with this physics-ML model is encouraging considering the simplified nature of boundary forcings and the low computational costs (Supplementary Table 1).

This study has several caveats related to the comparison with operational prediction models and the simplified boundary forcings. Since the TC data of most operational climate prediction models are not publicly accessible, we were unable to comprehensively compare our results with state-of-the-art models and evaluate potential differences in prediction skills and computation costs (Supplementary Materials). The simplified boundary forcings used in this study rely on the persistence of anomalies and can be less reasonable for other initialization time. We speculate that more realistic representations of boundary forcings or the inclusion of coupled climate processes (e.g., land-atmosphere coupling) may help NeuralGCM to accomplish more skillful predictions of TC activity (e.g., Zhang et al. 2021) and other aspects of the Earth system (e.g., Yeager *et al* 2022). Such development can be accomplished by coupling NeuralGCM with statistical models, ML emulators, or other hybrid models of the ocean and other Earth system components.

Contributing to the rapidly evolving field of the ML-based climate modeling, this study demonstrates a practical application of the NeuralGCM and provides valuable insights for future model development. Our hindcast experiments with simplified boundary forcings shows that the NeuralGCM can represent the atmospheric responses to boundary forcings (e.g., SST) that are critical for the subseasonal-to-seasonal prediction. These experiments establish a performance baseline against which future model iterations can be benchmarked. Furthermore, our results suggest that the NeuralGCM holds significant potential as a foundation for developing a computationally affordable system for seamless subseasonal-to-seasonal prediction. Nevertheless, the evaluation also underscores some challenges of applying the current ML atmospheric models, including limitations inherited from training datasets and the lack of coupling among key climate system components. Recognizing that physical GCMs required decades of refinement to achieve milestones like simulating realistic TC activity (Manabe *et al* 1970, Roberts *et al* 2020, Zhao *et al* 2009), patience and continued effort are warranted despite recent breakthroughs in ML modeling efforts. We expect intensified collaboration among ML and physical science communities to alleviate many of the identified issues, ultimately accelerating the transformation of climate model development and applications.

## Acknowledgment

G.Z is supported by the faculty development fund of the University of Illinois at Urbana-Champaign, and the faculty fellowship of the Office of Risk Management and Insurance Research (ORMIR) of Gies Business School, and the U.S. National Science Foundation awards AGS-2327959 and RISE-2530555. The authors thank Sarah Henry and Drs. Zhuo Wang, Stephan Hoyer, and Dmitrii Kochkov for stimulating discussions about evaluating and refining the NeuralGCM simulations. GZ thanks the organizers of the Rossbypalooza workshop at the University of Chicago for providing a welcoming environment for attendees.

## Data Availability

The ERA5 dataset is accessible via the Climate Data Store of the Copernicus Climate Change Service (https://doi.org/10.24381/cds.6860a573). The IBTrACS dataset is archived by the U.S. National Centers for Environmental Information (https://www.ncei.noaa.gov/products/international-best-track-archive). The NeuralGCM code is available at GitHub (https://github.com/google-research/neuralgcm) under the Apache 2.0 license. The code used to generate the plots in this study is available the Zenodo repository (doi: https://doi.org/10.5281/zenodo.15319941).

Supplementary Materials for

**Advancing Seasonal Prediction of Tropical Cyclone Activity with a Hybrid AI-Physics Climate Model**

Gan Zhang[1*], Megha Rao[1], Janni Yuval[2], Ming Zhao[3]

[1] Department of Climate, Meteorology, and Atmospheric Sciences, University of Illinois at Urbana-Champaign

[2] Google Research

[3] Geophysical Fluid Dynamics Laboratory, National Oceanic and Atmospheric Administration

File Content

    Supplementary Text

    Supplementary Figures 1–16

    Supplementary Table    1

**Technical Overview of NeuralGCM and Hindcast Experiments**

NeuralGCM is a differentiable, hybrid atmospheric model that integrates a traditional atmospheric dynamical core with machine-learned parameterizations of unresolved physics. The entire framework is implemented in JAX, a high-performance Python library that enables automatic differentiation, which is crucial for gradient-based training. This technical description of NeuralGCM closely follows Kochkov et al. (2024) and serves as a self-contained overview of the model. Additional technical details of the model are available in Kochkov et al. (2024) and its supplementary materials.

*a. Model Architecture and Components*

- Dynamical Core: The model's foundation is a differentiable dynamical core that solves the hydrostatic primitive equations with moisture. It employs a pseudo-spectral method for horizontal discretization and a vertical sigma-coordinate system, which is well-suited for simulating large-scale atmospheric dynamics. The atmospheric core evolves seven prognostic variables: vorticity, divergence, temperature, surface pressure, and the specific content of three water species (humidity, cloud ice, and cloud liquid).
- Learned Physics Module: Sub-grid scale physical processes (e.g., convection), which are not resolved by the dynamical core, are parameterized by a neural network. This module follows the single-column approach common in General Circulation Models (GCMs), where the physical tendencies for a given atmospheric column are predicted based solely on the state of that column. The architecture is a fully connected neural network with residual connections, and a single set of weights is shared across all columns of the global grid. Inputs to this network include the local prognostic variables, boundary conditions

(sea-surface temperature, sea-ice concentration), incident solar radiation, learned spatial embeddings, and the horizontal gradients of prognostic variables. All inputs are standardized, and the network outputs are scaled tendencies for the prognostic variables.

- Encoder-Decoder for Data Interfacing: NeuralGCM uses an encoder-decoder system to bridge the gap between its native sigma-coordinate system and the pressure-level coordinates used by the training dataset (i.e., ERA5). These components perform linear interpolation but are augmented with learned neural network corrections. This learned component is critical for model performance, as it reduces the initialization shock and the resulting contamination of forecasts by spurious gravity waves.

b. Simulation Workflow

The encoder ingests initial conditions from ERA5 data on pressure levels and transforms them into the model's sigma-coordinate system. After the initialization, the model state is advanced in time using an implicit-explicit Ordinary Differential Equation (ODE) solver. In each step, the solver uses tendencies calculated by both the dynamical core and the learned physics module. To optimize performance, the computationally expensive physics tendencies are calculated less frequently (e.g., every 30 minutes of simulation time) and held constant over multiple, shorter time steps of the dynamical core, which are constrained by the Courant–Friedrichs–Lewy (CFL) condition. After the final forecast time is reached, the decoder converts the model's predictions from sigma coordinates back to standard pressure levels for analysis and evaluation. Sample simulation code is available at the model webpage (https://neuralgcm.readthedocs.io/en/latest/inference_demo.html).

c. Training Methodologies

- Deterministic Model: The deterministic version is trained with a composite loss function designed to optimize three distinct objectives. A modified Mean Squared Error (MSE) loss term progressively filters out small-scale features at longer lead times. Another loss term encourages the wavenumber spectrum of the predicted fields to match the climatological spectrum of the training data. A third loss term penalizes the mean error in each spherical harmonic coefficient and discourages systematic drift. This combined loss function enables the model to produce stable, long-term climate simulations. The final model is also subject to a fine-tuning stage focused on the decoder component.

- **Stochastic Model:** The stochastic version is designed to generate probabilistic forecasts by accounting for model uncertainty. Randomness is injected into the model by adding spatially and temporally correlated Gaussian noise as an input to the learned physics and encoder modules. It is trained using the Continuous Ranked Probability Score (CRPS), a metric that rewards both forecast accuracy and a well-calibrated ensemble spread. The CRPS loss is computed in both grid space and spectral space to ensure probabilistic skill across different scales.

d. Hindcast Experiments

The hindcast experiments for 5-month predictions follow the simulation workflow described earlier and incorporate three modifications. We first modify the boundary forcing input as described in the main text. The combined fields of SST and sea ice anomalies present at the initialization time and the smoothed daily climatology serve as the time-varying boundary forcing. To improve the simulation stability, we also enable a constraint that fixes the global mean log surface pressure. This modification has been documented (https://neuralgcm.readthedocs.io/en/latest/checkpoint_modifications.html) and does not need re-training for fine-tuning of the publicly available NeuralGCM versions. Lastly, we introduce initial condition perturbations to the deterministic runs by modifying the interpolation of the ERA5 input to model levels. As described by Section 2.2, this interpolation involves an encoder and a learned correction, and this correction is perturbed to generate initial conditions for the ensemble prediction with the deterministic model.

**Considerations for Model Evaluation**

This study produces and evaluates the NeuralGCM's performance with simplified boundary conditions for seasonal prediction tasks. While this configuration provides a valuable baseline, it represents a lower-bound estimate of the model's potential skill due to several limitations, primarily stemming from the simplified boundary forcing used to generate the hindcast and the biases of inherited from the training dataset (i.e., ERA5) in representing tropical cyclones. Direct comparisons with operational seasonal prediction models are challenging due to data accessibility. For example, the computation costs of operational physical models are rarely disclosed, making it difficult to quantify the advantage of the NeuralGCM configuration. Additionally, the specific variables and temporal resolution required for a fair comparison of model skill are often not readily accessible due to various factors. For instance, the 6-hourly surface pressure and upper-air fields of operational physical models are crucial for detailed verification and tropical cyclone tracking, but they are not consistently archived or distributed by public service. Furthermore, the NeuralGCM version used by this study does not directly output precipitation or surface temperature, two key variables for many forecast applications. Lastly, the boundary forcings and the representation of coupled processes differ substantially between this NeuralGCM configuration and the operational prediction systems. This issue makes it hard to attribute the sources of skill differences, limiting scientific insights that can be yielded by direct comparisons.

With those caveats in mind, we include anomaly correlation coefficients of an operational seasonal prediction system (Johnson et al. 2019) and the observation for a set of available variables (Supplementary Figures 8–10) to facilitate preliminary comparisons of prediction skill. The ECMWF seasonal forecasts (SEAS5) are produced every month with a 51-member ensemble at a horizontal resolution of ~36 km. The SEAS5 model incorporates coupled physical models that represent the atmosphere, ocean, and sea ice processes (Johnson et al. 2019). Despite being more computationally intensive, the representation of ocean and sea ice by SEAS5 is much more realistic than the simplified boundary forcings used by our NeuralGCM configuration. A subset of the SEAS5 prediction data is distributed via data interfaces (e.g., wind) or in graphic format (e.g., TC metrics). The hindcast of the SEAS5 version 5.1 is available from 1981, but only 25 members are available in the years before 2017. To facilitate comparisons with the NeuralGCM hindcasts, we randomly select 20 members of the SEAS5 ensemble initialized on July 1 of 1990-2019.

Our preliminary comparison suggests that SEAS5 delivers more skillful predictions of the environmental variables (Supplementary Figures 8–10) than our NeuralGCM configuration (Supplementary Figures 4, 6, and 7), especially at longer time leads. In this preliminary evaluation, the regions with relatively high prediction skill are consistent in the SEAS5 and NeuralGCM predictions at each forecast lead. This suggests the behaviors of our NeuralGCM configuration are generally consistent with physical processes represented by the SEAS5, despite some artifacts in the NeuralGCM hindcast (see next section). Whether the skill difference in predicting environmental variables leads to differences in TC prediction remains to be investigated. A comparison of the TC prediction is difficult since the SEAS5 TC data is only available period starting from 1992, and the publicly accessible graphic format includes additional bias corrections of TC metrics. Based on the publicly available, bias-corrected TC data from the SEAS5

([https://charts.ecmwf.int/products/seasonal_system5_tstorm_verification](https://charts.ecmwf.int/products/seasonal_system5_tstorm_verification)), we estimate that the skill of the SEAS5 (with bias corrections) and NeuralGCM in predicting basin-wide TC counts and ACE are comparable. A more comprehensive evaluation needs to involve collaboration with developers of operational seasonal prediction models to establish a standardized comparison framework, including common output variables and consistent data processing (e.g., detrending, see Wulff et al. 2022), facilitating a more rigorous and insightful assessment of the model capabilities.

**Comparisons of NeuralGCM Hindcasts with Deterministic and Stochastic Physics**

These hindcasts show similar computational efficiency and skills in predicting TC metrics, but their simulation stability and biases appear to have notable differences. We observed minor differences in the simulation speed between the deterministic and stochastic models (Supplementary Table 1). For the 20-member 160-day simulations in our test environment, the 1.4-degree NeuralGCM with stochastic physics uses about 5% more time than its counterpart with deterministic physics. The deterministic and stochastic models perform similarly in predicting the large-scale environment (not shown) and TC metrics, including TC frequency and accumulated cyclone energy (Table 1, Figure 4, and Supplementary Figure 12). In terms of model instability, the hindcast simulations with stochastic physics have a higher chance to encounter issues with spurious waves in the tropics (main text and Supplementary Figures 1-2). The hindcast simulations with deterministic physics appear more likely to generate large biases in polar regions, which contribute to large values of normalized root-mean-squared errors (e.g., Supplementary Figures 4, 6, and 7). Such large polar biases are not apparent in the hindcast simulations with stochastic physics. More detailed comparisons between the models with deterministic and stochastic physics will be conducted after an update of the NeuralGCM.

**Performance with Environmental Controls, Notable Years, and Subseasonal Activity**

The TC-environment relationship in the NeuralGCM hindcast and the observation are generally consistent, especially in the North Atlantic and the Northwest Pacific (Supplementary Figure 14). The analysis here focuses on the surface pressure and the vertical wind shear that show relatively strong correlations with basin-wide TC counts. For the TC frequency in the North Atlantic, the correlations with the surface pressure and the zonal wind shear are strong, and the NeuralGCM hindcasts and the observation show consistent patterns in the tropics. Relatively strong correlations and observation-consistent patterns are also present for the TC frequency in the Northeast Pacific. However, the correlation patterns are less consistent with the observation for the Northwest Pacific, including the nearby region of Southeast Asia. For example, the correlation with the vertical wind shear is over strong near Southeast Asia in the NeuralGCM hindcast. For the TC frequency in the North Indian Ocean, the TC-environment correlation is much weaker than its counterpart of the other basins. This weak correlation, as well as the low skill in predicting local environmental variables (e.g., Supplementary Figure 7), is consistent with the relatively poor skill in predicting TC activity in the North Indian Ocean (Figure 4).

The NeuralGCM hindcasts of the 2018–2023 seasons, which is not used to train the deterministic model, also show some skill in the North Atlantic and the Northeast Pacific (Figure 4). However, the NeuralGCM hindcast tends to under-predict TC counts in extreme seasons, and we thus use two extreme Atlantic hurricane seasons (2005 and 2020) to examine potential causes

(Supplementary Figure 15). In both seasons, NeuralGCM manages to predict the large-scale patterns in the tropics, which have strong associations with TC frequency (Supplementary Figure 14). The prediction in the extratropics show some notable disagreement with the observation (e.g., 500-hPa geopotential in 2005). However, extratropical impacts appear less important for TC activity in the deep tropics (Zhang et al. 2019). A comparison of the simulated and observed TC tracks suggests the underprediction of TC frequency is related to the absence of activity in the Caribbeans and the Gulf of Mexico. This absence is particularly notable for the 2020 hurricane season and makes the hindcasts miss about a half of TCs. Moreover, the hindcast misses the region's low-level pressure anomalies in the 2020 observation (Supplementary Figure 15g and 15h), consistent with the relatively poor long-term skill (Supplementary Figure 5). Since NerualGCM forced with observed SST successfully simulated TC activity in the Caribbeans and the Gulf of Mexico in 2020 (Extended Figure 7 of Kochkov et al. 2024), this regional underprediction of TC activity in our hindcast appears related to the simplified boundary forcing. Improved SST prediction and more realistic atmosphere-ocean coupling might help alleviate such regional biases in the North Atlantic and other basins.

Analyses of the NeuralGCM hindcasts also show promise in simulating and predicting monthly TC activity (Supplementary Figure 16). The monthly climatology of TC frequency in the observation and the NeuralGCM hindcast are generally consistent, though phase biases related to depressed late-season activity are notable with NeuralGCM (also See Fig. 3). The prediction skill of TC counts, as indicated by the correlation coefficient, is the highest in July (Month 1) and decays in later months. The skill decay across individual months is not monotonic (e.g., Supplementary Figure 16b), suggesting a larger number of predictions that sample more initialization time are needed to robustly evaluate the subseasonal prediction skill. Consistent with the seasonal prediction skill, the skill in predicting monthly activity is the highest in the North Atlantic and the Northeast Pacific. The low subseasonal skill in the Northwest Pacific and the North Indian Ocean might be related to seasonal-scale issues discussed in earlier this section.

**Additional References**

- Wulff, C.O., Vitart, F. & Domeisen, D.I.V.(2022) Influence of trends on subseasonal temperature prediction skill. *Quart J Royal Meteoro Soc*, 148, 1280–1299.
- Zhang, G., Knutson, T. R., & Garner, S. T. (2019). Impacts of extratropical weather perturbations on tropical cyclone activity: Idealized sensitivity experiments with a regional atmospheric model. *Geophysical Research Letters*, 46, 14052–14062.

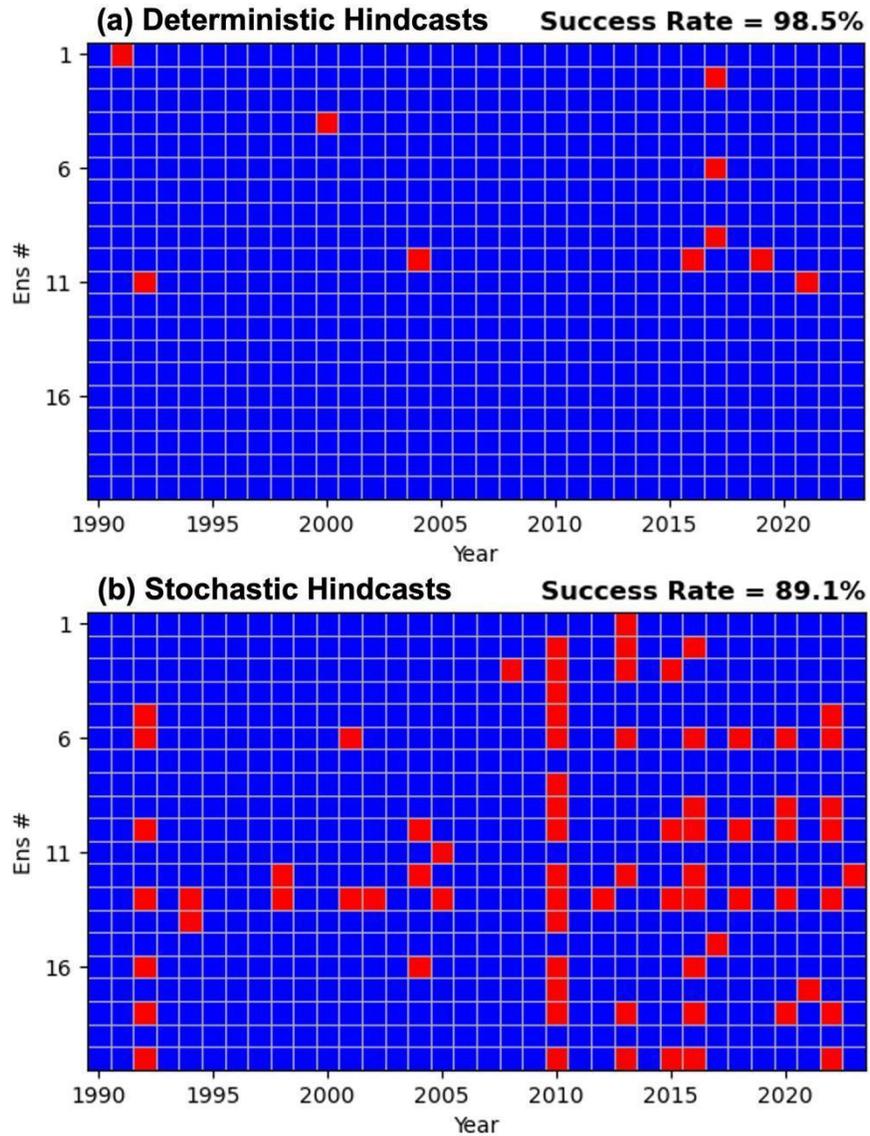

Supplementary Figure 1 *An overview of rollout instability in the NeuralGCM hindcasts. (a) Hindcasts with the deterministic physics. (b) Hindcasts with the stochastic model physics. The instability cases are flagged with red coloring. The percentage of simulations that passed the stability test is denoted in the top right of subplots.*

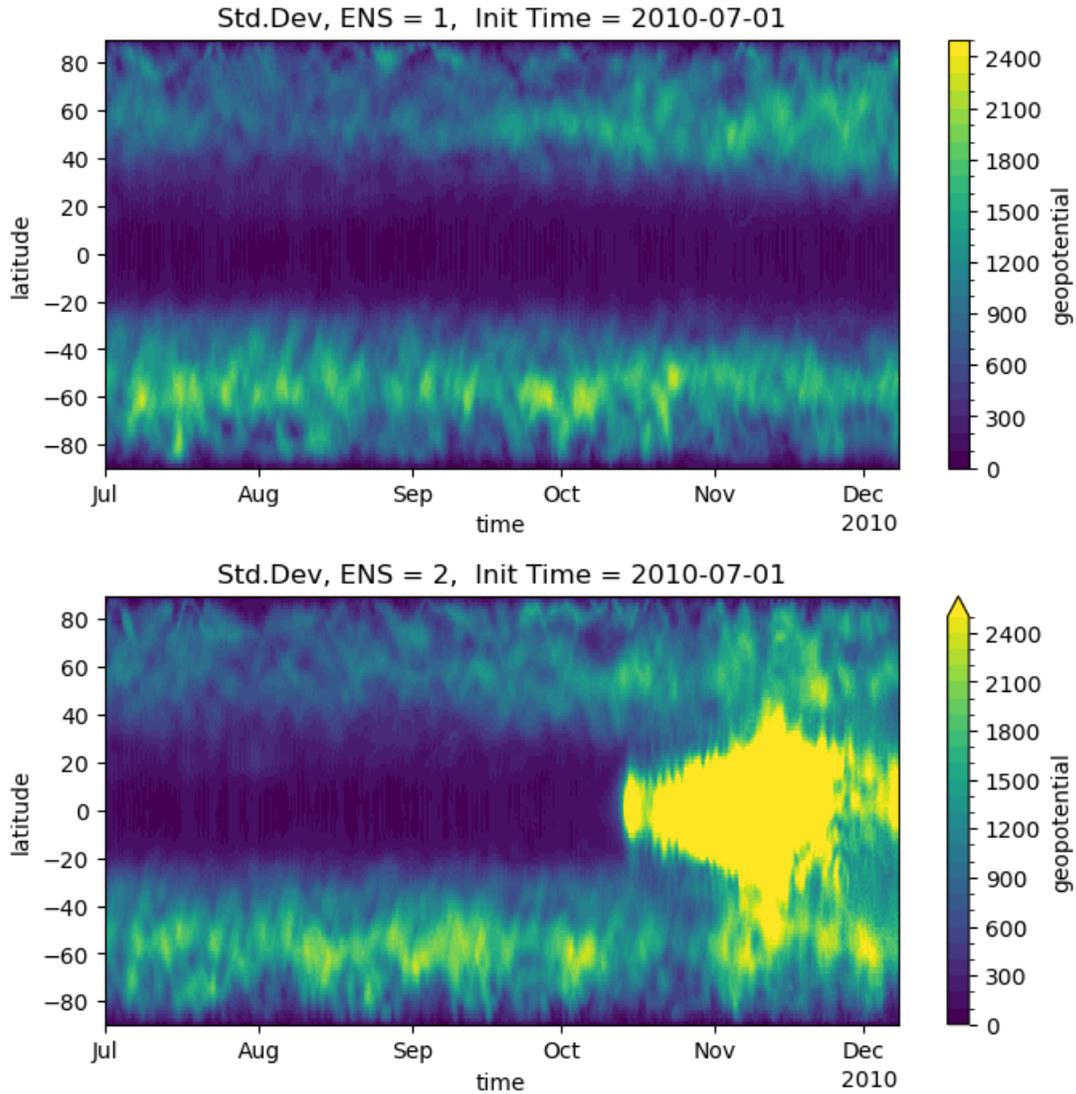

Supplementary Figure 2 *Evolution of stable (top) and unstable (bottom) prediction rollouts with stochastic model physics. The field is the standard deviation of the 500-hPa geopotential ($m^2\ s^{-2}$) in the zonal direction. Spurious waves appear in the unstable simulation around mid-October. Both simulations were initialized on July 1, 2010 with slightly different random seeding.*

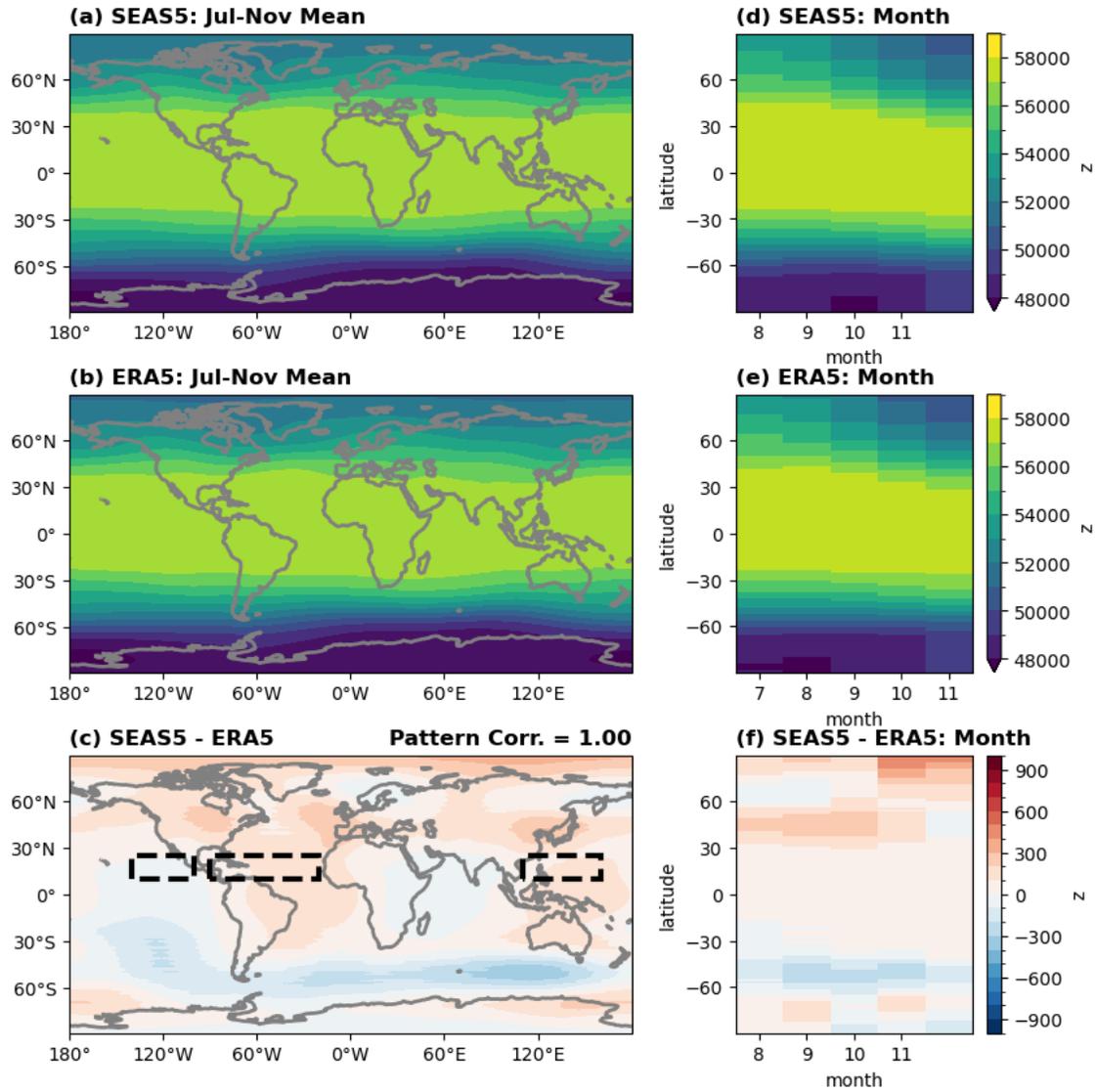

Supplementary Figure 3 *The climatology of 500-hPa geopotential ($m^2\ s^{-2}$) of the ECMWF's fifth generation seasonal forecast system (SEAS5). The data of SEAS5 (version 5.1) of 1990-2019 are regridded to the NeuralGCM grid. The original SEAS5 data has 25 ensemble members for simulations initialized in 1990-2016 and 51 for simulations initialized in 2017-2019. The analyses only consider the first 20 ensemble members to ensure the ensemble size is consistent with the NeuralGCM hindcasts. The plotting settings are the same as Figure 1.*

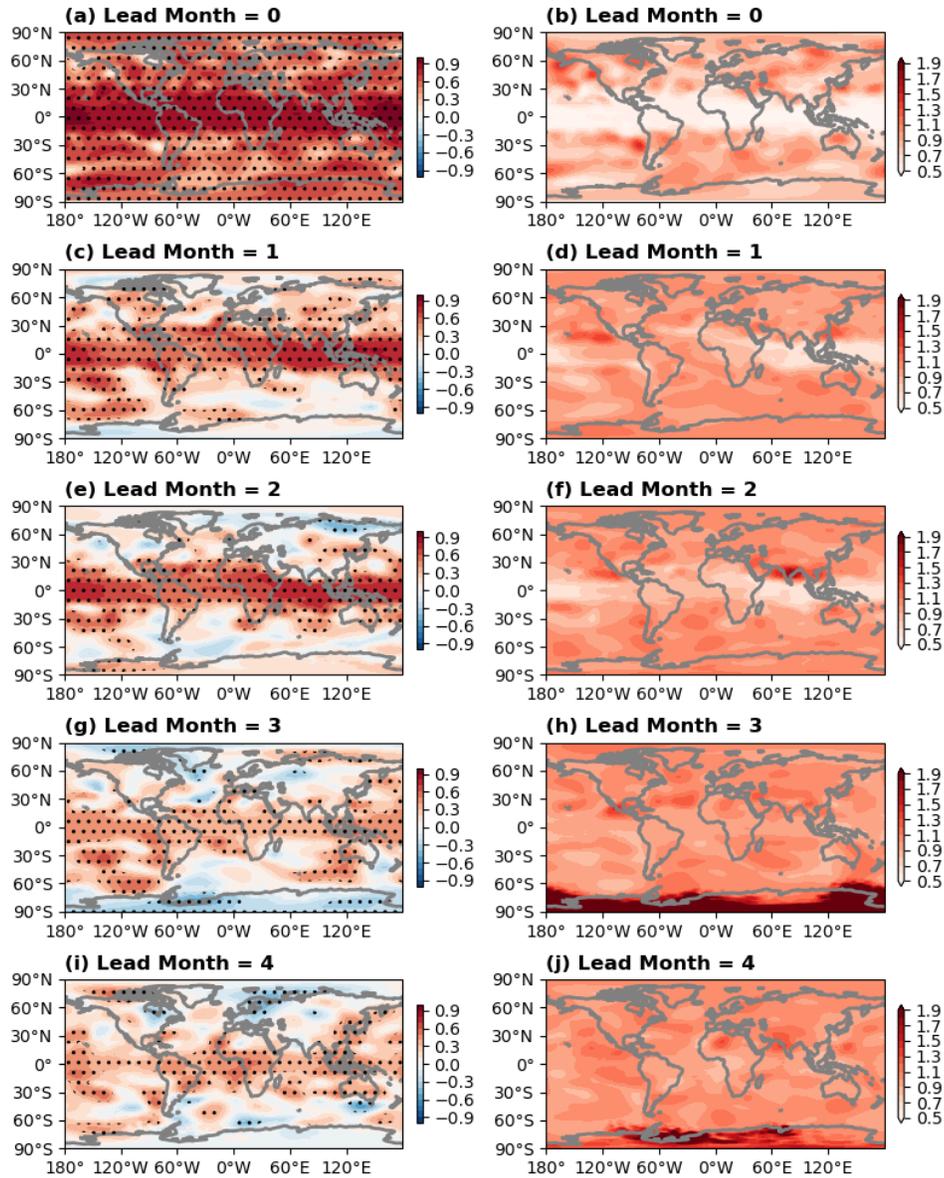

Supplementary Figure 4 *Skill metrics of the NeuralGCM hindcasts for the 500-hPa geopotential ($m^2\ s^{-2}$)*. (a) Anomaly correlation coefficients between the NeuralGCM hindcasts and the ERA5 reference in Lead Month 0 (July). (b) Normalized root mean squared error (RMSE) of the 500-hPa ($m^2\ s^{-2}$) of the NeuralGCM hindcasts relative to the ERA5 reference in Lead Month 0. (c, e, g, i) Same as (a), but for Lead Month 1–4 (August–November), respectively. (d, f, h, j) Same as (b), but for Lead Month 1–4 (August–November), respectively. The input variables were detrended using a linear least-squared fitting. The stippling in the left column indicates parts above the 90% confidence level.

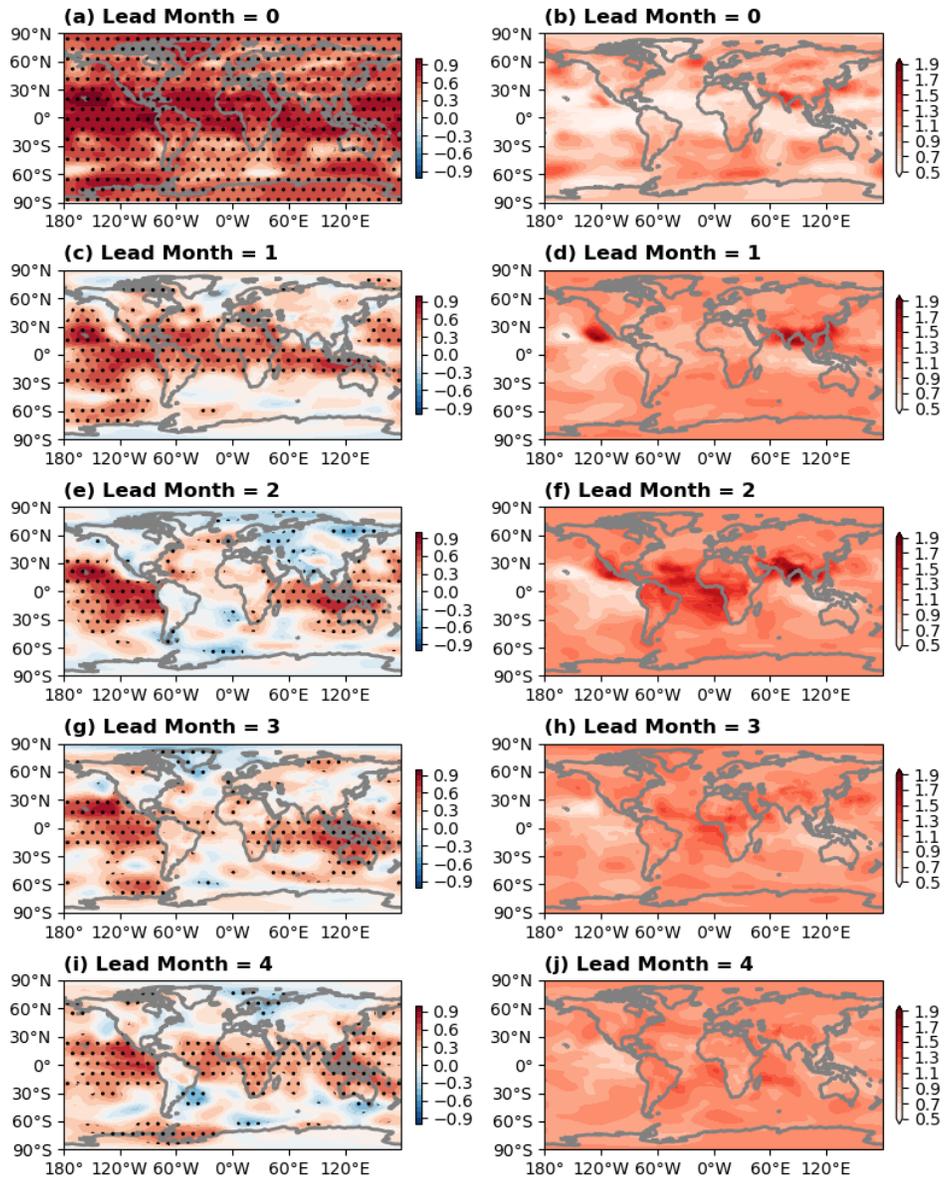

Supplementary Figure 5 *Skill metrics of the NeuralGCM hindcasts for the surface pressure (hPa). The other settings are the same as in Supplementary Figure 4. The input variables were detrended using a linear least-squared fitting.*

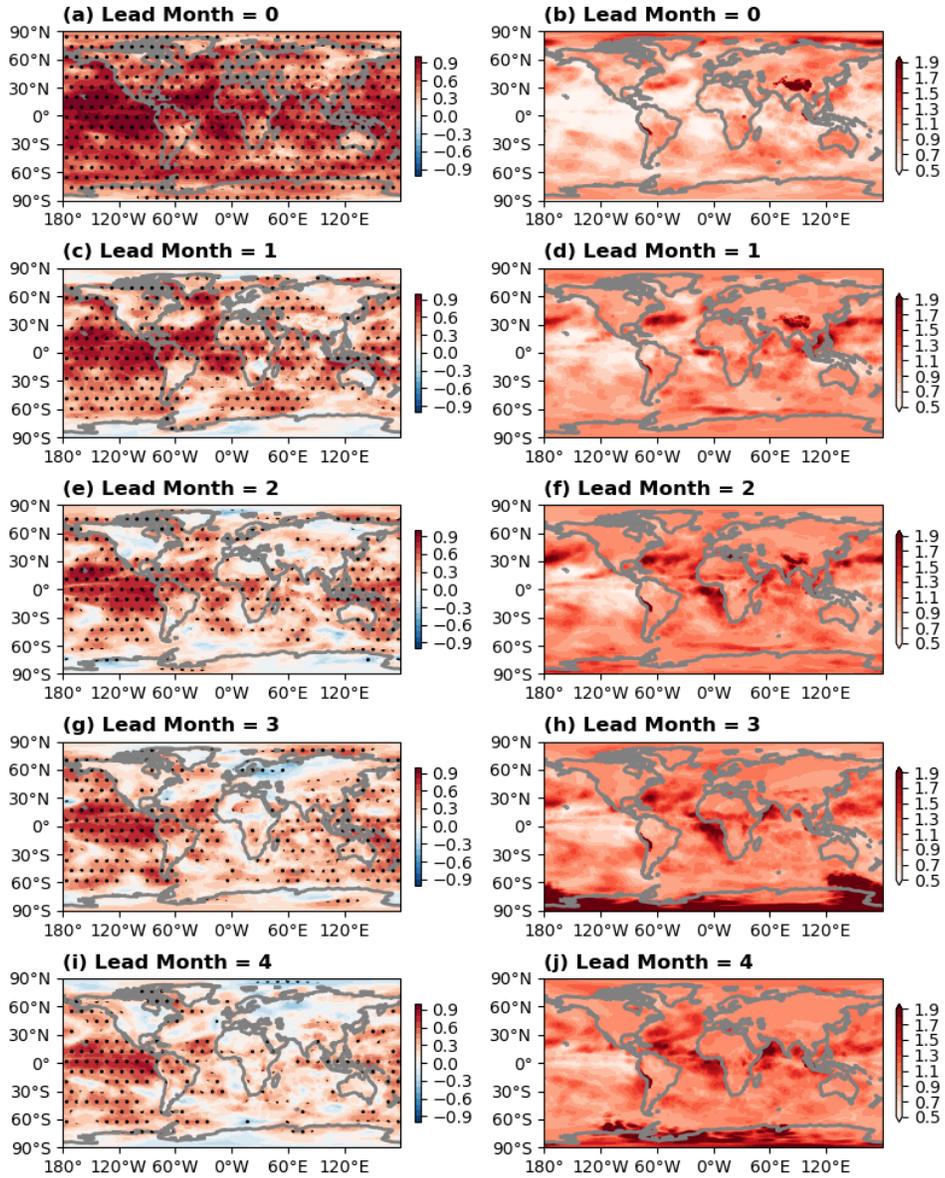

Supplementary Figure 6 *Skill metrics of the NeuralGCM hindcasts for the 1000-hPa air temperature (K). The other settings are the same as in Supplementary Figure 4. The input variables were detrended using a linear least-squared fitting.*

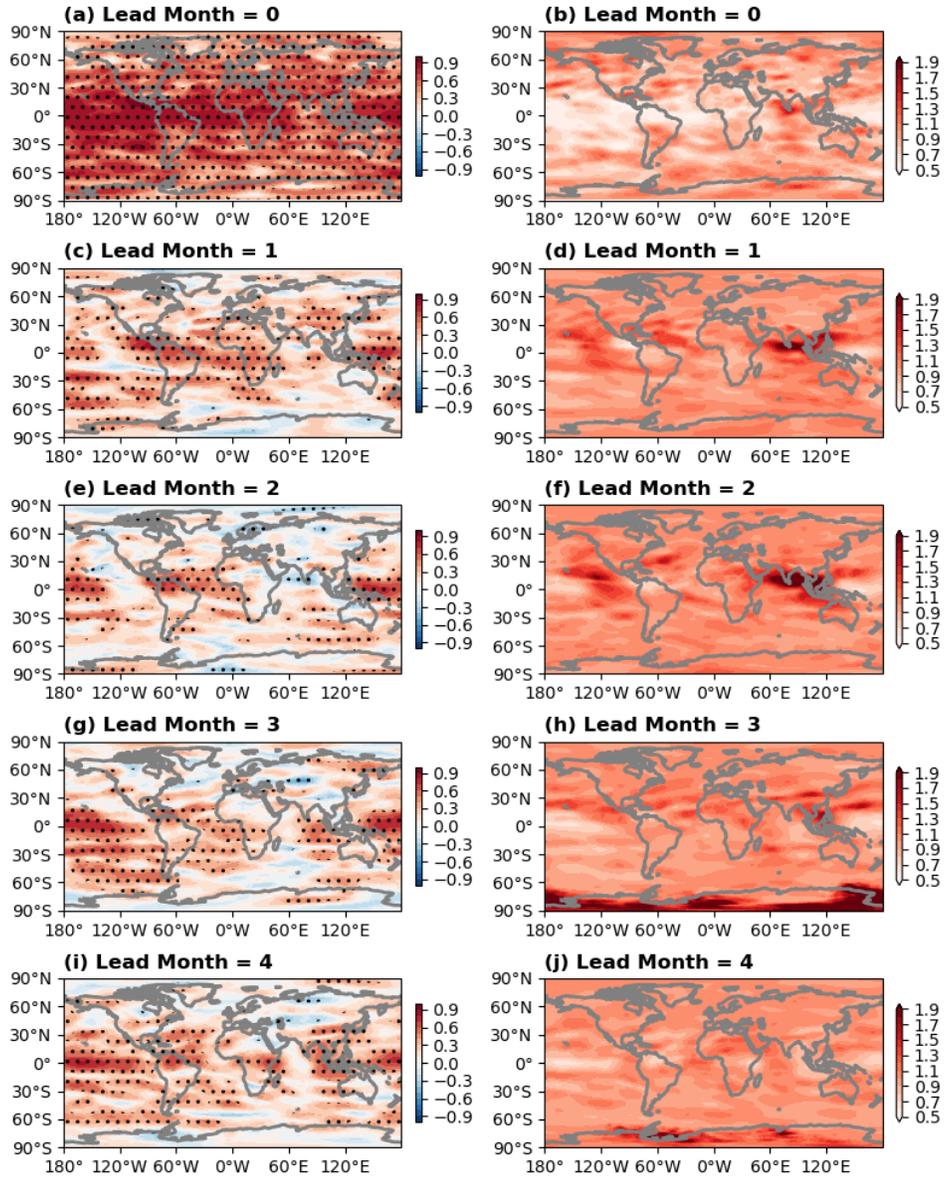

Supplementary Figure 7 *Skill metrics of the NeuralGCM hindcasts for the zonal wind shear (200 hPa - 850 hPa; unit: m s$^{-1}$). The other settings are the same as in Supplementary Figure 4. The input variables were detrended using a linear least-squared fitting.*

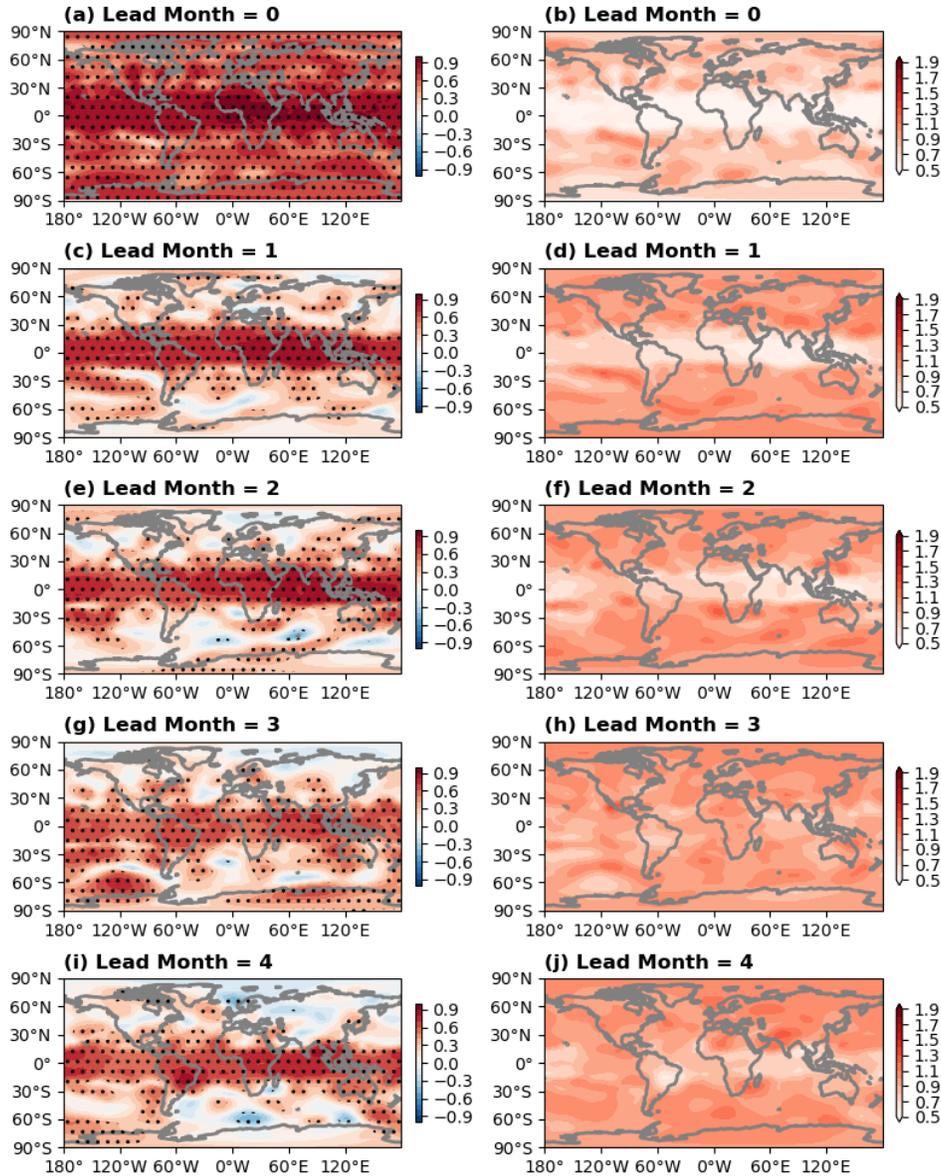

Supplementary Figure 8 *Skill metrics of the ECWMF SEAS5 for the 500-hPa geopotential ($m^2$ $s^{-2}$). The data of SEAS5 (version 5.1) of 1990-2019 are regridded to the NeuralGCM grid. The analyses only consider the first 20 ensemble members to ensure the ensemble size is consistent with the NeuralGCM hindcasts. The input variables were detrended using a linear least-squared fitting. The other plotting settings are the same as Supplementary Figure 4.*

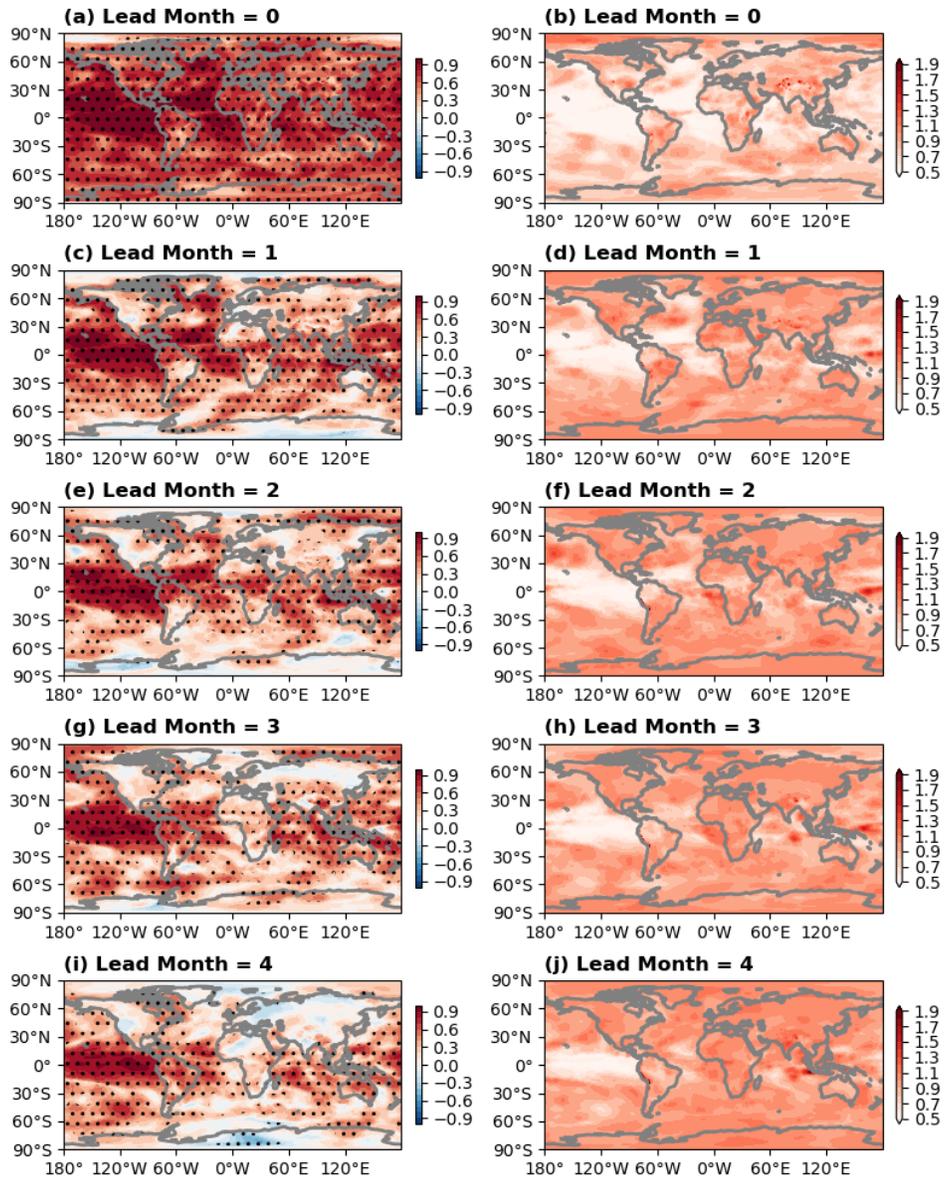

Supplementary Figure 9 *Skill metrics of the ECWMF SEAS5 for the 1000-hPa temperature (K) The other plotting settings are the same as Supplementary Figure 4. The input variables were detrended using a linear least-squared fitting.*

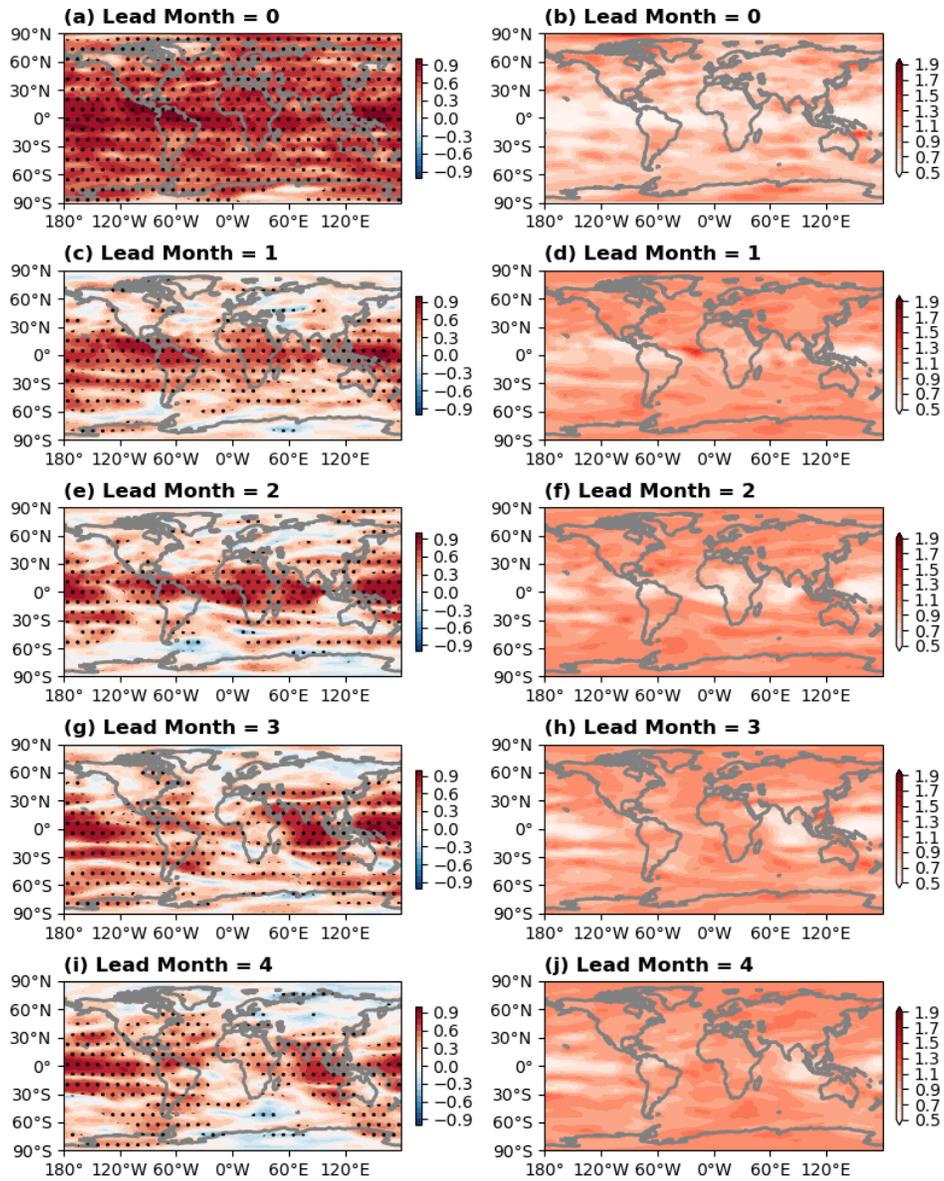

Supplementary Figure 10 *Skill metrics of the ECWMF SEAS5 for the 200-hPa and 850-hPa zonal wind shear (m s$^{-1}$). All the other settings are the same as Supplementary Figure 4. The input variables were detrended using a linear least-squared fitting.*

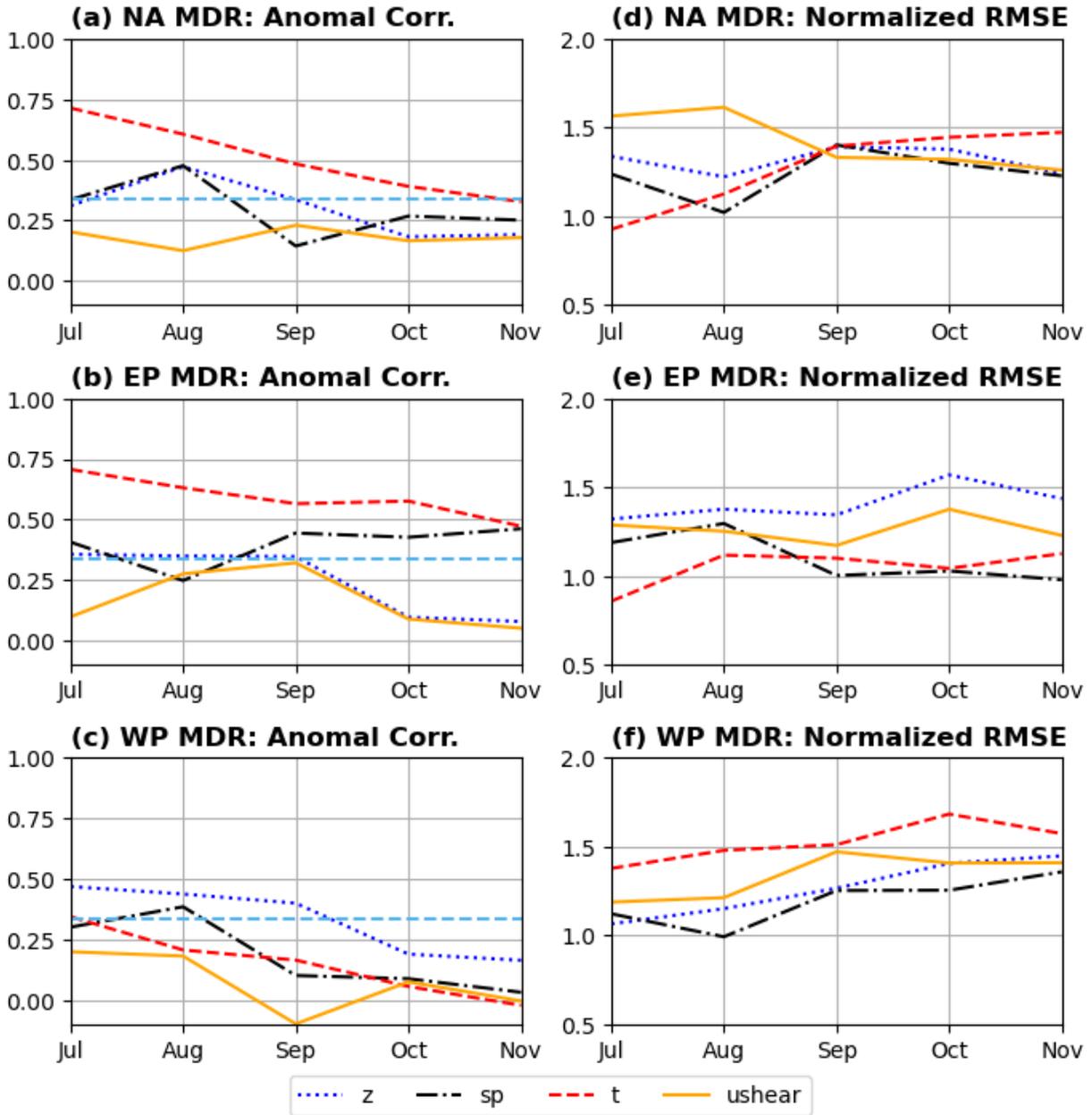

Supplementary Figure 11 *Prediction skills of environment variables in the Main Development Regions (MDRs) by persisting the June mean anomalies. The input variables were detrended using a linear least-squared fitting. The y-axis range in the right column is adjusted to better display values. The other settings are the same as Figure 2.*

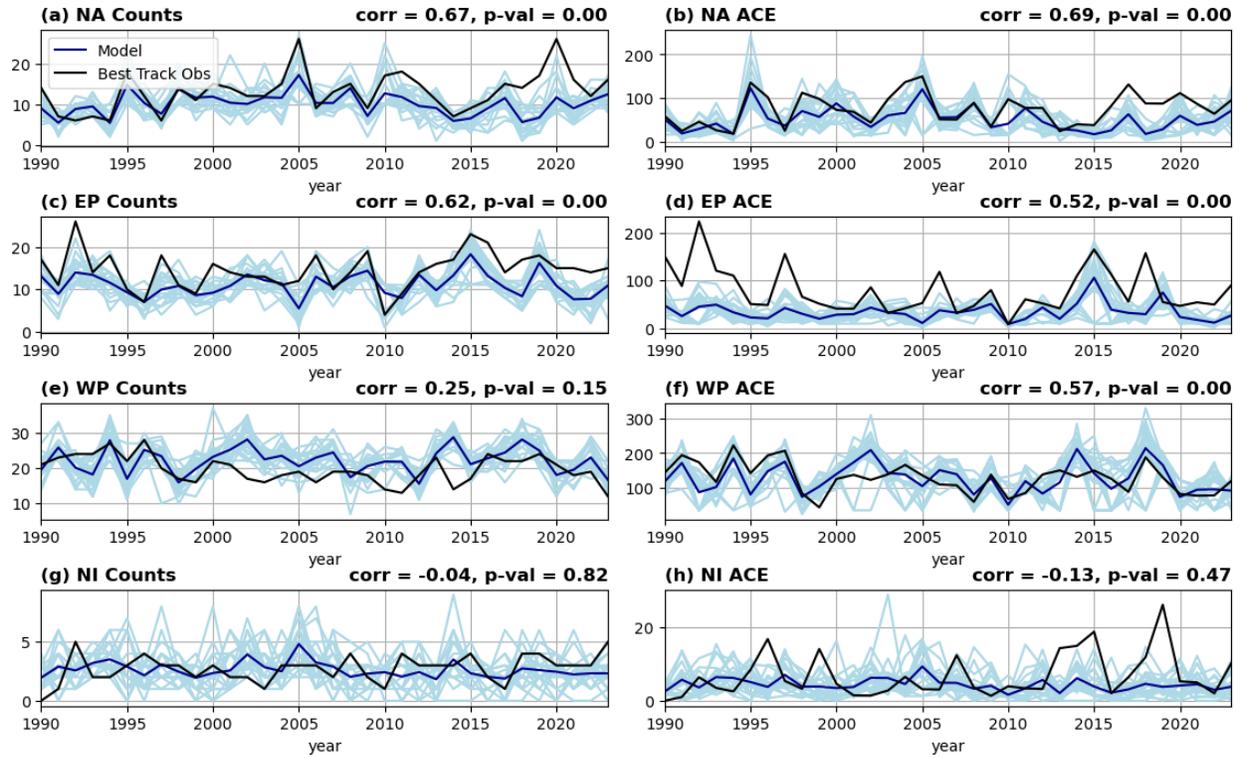

Supplementary Figure 12 *Predictions and observations of the basin-wide TC counts (1990-2023) and the accumulated cyclone energy (ACE; unit: $10^{-4}$ knot$^2$) using the NeuralGCM with stochastic model physics. The other settings are identical to Figure 4.*

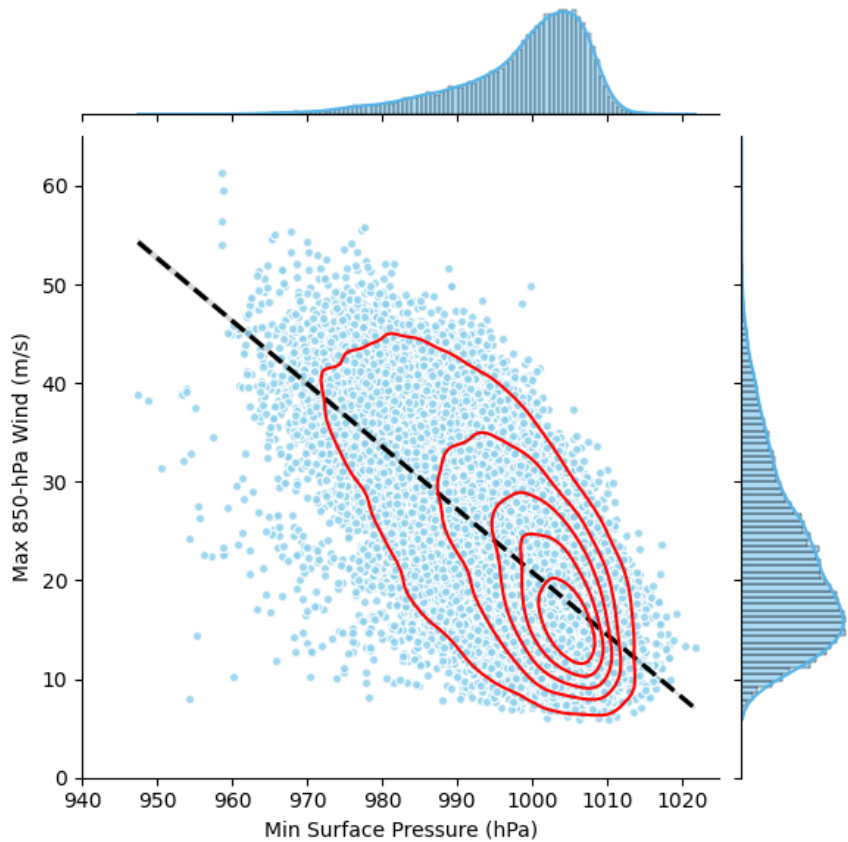

Supplementary Figure 13 (a) The relationship between the surface pressure (hPa) and the 850-hPa maximum wind speed (m s$^{-1}$) in the NeuralGCM hindcasts with the deterministic model physics. The blue dots show the individual data points. The red contours show the kernel density estimate of samples. The histograms on the right and the top show the distributions of the maximum 850-hPa wind speed and the minimum surface pressure. The surface pressure and wind speed of TCs are inversely correlated.

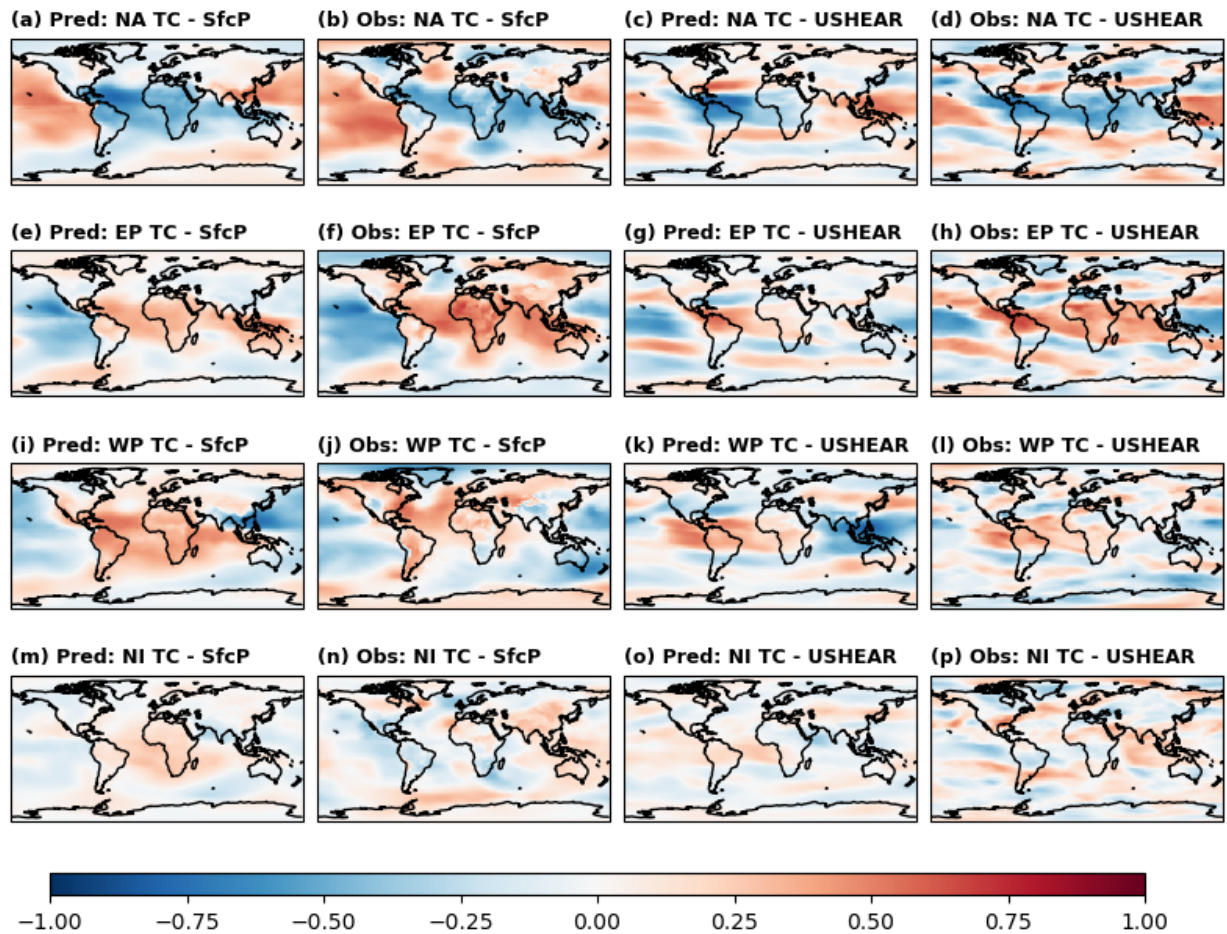

Supplementary Figure 14 The relationship between the large-scale atmospheric environment and TC counts during July–November of 1990–2023. (a) Correlation between North Atlantic TC counts and surface pressure in the NeuralGCM hindcast. (b) Same as (a), but for the observational data (i.e. best track and ERA5). (c) Same as (a), but for the vertical shear of 200-hPa and 850 hPa zonal wind. (d) Same as (b), but for the vertical wind shear. (e–h), (i–l), and (m–p), are the same as (a–d), but for the East Pacific, Northwest Pacific, and North Indian Ocean, respectively. The threshold of 95% confidence level for the Pearson correlation coefficients is approximately $\pm 0.34$ as determined with two-tailed Student's t-distribution (N=34).

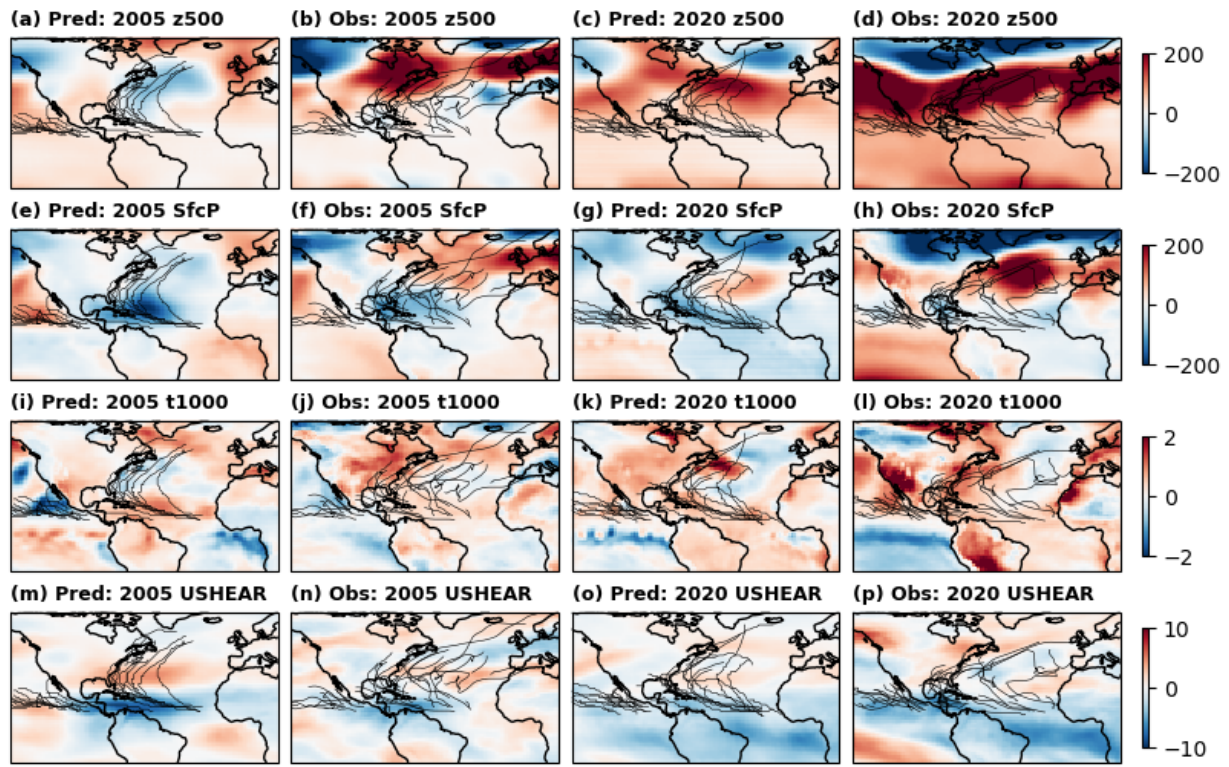

Supplementary Figure 15 The predictions and observations of the *large-scale atmospheric environment and TC counts during two extreme Atlantic hurricane seasons*. (a) The NeuralGCM prediction of 500-hPa geopotential height anomalies (shading; $m^{-2} s^{-2}$) and TC tracks (black lines) in July–November of 2005. The geopotential height is the 20-member ensemble mean, and the TC tracks are from a random ensemble member instead of all ensemble members for visual clarity. (b) Same as (a), but for the observation (i.e. best track and ERA5). (c)(d) Same as (a)(b), but for July–November of 2020. (e–h), (i–l), and (m–p), are the same as (a–d), but for the anomalies of surface pressure (Pa), 1000-hPa temperature (K), and the vertical shear of 200-hPa and 850 hPa zonal wind ($m\ s^{-1}$).

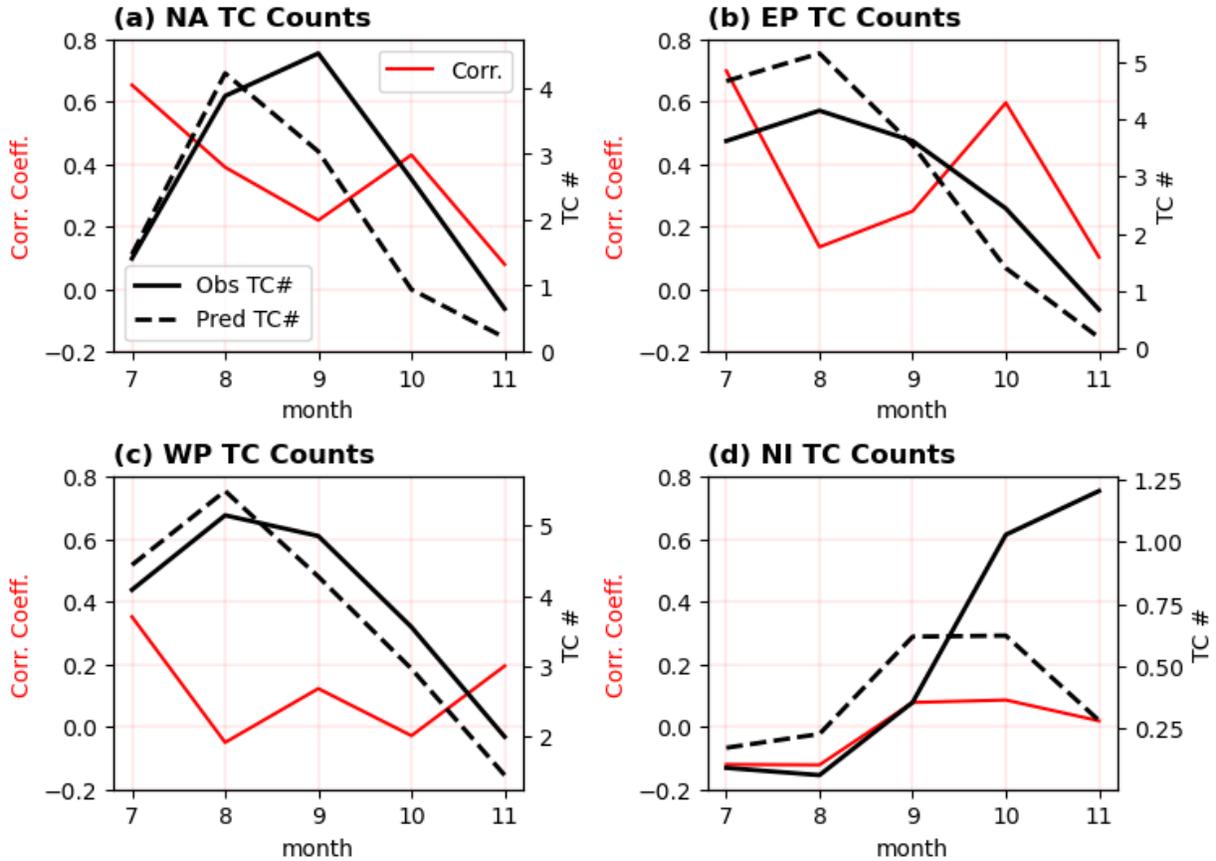

Supplementary Figure 16 *Monthly tropical cyclone counts and prediction skill (1990–2023). (a) Results of the North Atlantic basin. The black solid and dashed lines show the long-term means of TC counts of the observation and the Neural GCM hindcasts, respectively. The red line shows the correlations of year-to-year variations by month. (b)(c)(d), same as (a) but for the Northeast Pacific (EP), Northwest Pacific (WP), and North Indian Ocean (NI).*

*Supplementary Table 1 Computation time of conducting hindcast experiments using NeuralGCM (1.4-degree) with deterministic and stochastic physics. All the experiments are conducted on single Nvidia L40S GPUs (CUDA version = 12.3, Driver version = 545.23.08). Each experiment includes 20-member 160-day simulations initialized on July 1$^{st}$ of different years. The data is saved per 6 simulation hours. The computation time does not include the preparation of the boundary forcing files, which takes trivial time (<1 min) due to the simplicity of the persisting SST and sea ice anomalies. Additional performance data of NeuralGCM on Google Tensor Processing Units (TPUs) is available in Extended Data Table 1 of Kochkov et al. (2024).*

| | **Deterministic Physics** | | **Stochastic Physics** | |
|---|---|---|---|---|
| **Experiment ID** | **Total Wall-clock Time (min)** | **Wall-clock Time (min) per 100 Simulation Days** | **Total Wall-clock Time (min)** | **Wall-clock Time (min) per 100 Simulation Days** |
| **Run #1** | 239 | 7.5 | 253 | 7.9 |
| **Run #2** | 241 | 7.5 | 254 | 7.9 |
| **Run #3** | 244 | 7.6 | 255 | 8.0 |
| **Run #4** | 240 | 7.5 | 253 | 7.9 |